\documentclass[12pt,a4paper]{article}%

\usepackage{amsmath,amssymb,amsfonts}
\usepackage{calc}
\usepackage{caption2}
\usepackage{hyperref}
\usepackage[pdftex]{color,graphicx}

\numberwithin{equation}{section} 
\setcaptionmargin{1cm}
\newcommand{\hhref}[1]{\href{http://arxiv.org/abs/#1}{arXiv:#1}}

\graphicspath{{fig/}}

\begin{document}
\begin{titlepage}
\vskip 1.0cm
\begin{center}
{\Large \bf Higgs boson signals in $\lambda$SUSY\\
 with a Scale Invariant Superpotential  }
\vskip 1.0cm {\large  Enrico Bertuzzo, Marco Farina} \\[1cm]
{\it Scuola Normale Superiore and INFN, Piazza dei Cavalieri 7, 56126 Pisa, Italy} \\[5mm]
\vskip 1.0cm
\today
\end{center}

\begin{abstract}
We reconsider the $\lambda$SUSY model, in which the mass of the Higgs boson is raised already at tree level up to about
$200$ GeV via an additional
interaction in the superpotential between the Higgs bosons and a Singlet, focusing on a scale invariant superpotential.
After a detailed analysis of the allowed region in parameter space, which includes constraints coming from the absence of
spontaneous CP violation and of unrealistic minima, we study the scalar mass spectrum, the production rate and the decay modes
of the lightest scalar, finding that in general both the production rate and the Branching Ratio into gauge bosons can be
reduced with respect of those of the standard Higgs, causing it to be hardly detected in the first run of the LHC.
\end{abstract}
\end{titlepage}

\section{Introduction }
It is plausible that two years from now the LHC will have collected up to $5 fb^{-1}$ of data in its $7$ TeV running.
Among other interesting searches, the one focusing on the Higgs boson seems to be very promising, since the last projections show an
exclusion potential up to a mass of $600$ GeV for a scalar with the same couplings as the Standard Model (SM) Higgs
\cite{CMS}. Nonetheless, it makes sense to ask: what if the LHC does not find anything? Does it means
that the Higgs paradigm for Electroweak Symmetry Breaking (EWSB) is ruled out?\\
Of course the answer is negative, since in many motivated extensions of the SM the lightest scalar of the theory can have very
different properties with respect to the SM Higgs particle (different mass and different couplings, affecting both
the production cross section and the decay modes).\\
In this work we focus on a particular supersymmetric framework, named $\lambda$SUSY \cite{lambdaSUSY}, in which one
additional
coupling between the two Higgs bosons and a singlet $\hat{S}$ of the SM gauge group,
$\Delta W= \lambda \hat{S} \hat{H}_1 \hat{H}_2$,
is added to the superpotential in order
to raise the Higgs boson mass above the LEP II experimental limit \cite{LEPHIGGS} without relying on radiative corrections. A large coupling $\lambda$ at low energies allows to increase the Higgs
boson mass already at tree level, in such a way that the well known problem of the MSSM Higgs boson mass with the LEP
lower bound is easily satisfied and also naturalness is improved \cite{Barbieri:2010pd}.\\
In this work we analyse a different superpotential with respect to the one studied previously in
\cite{lambdaSUSY}, namely the NMSSM scale invariant superpotential $W= \lambda \hat{S} \hat{H}_1 \hat{H}_2 + (k/3) \hat{S}$,
which is particularly interesting since it allows to solve the so called $\mu$ problem
of the MSSM. Indeed in this case the charged higgsino mass term results $\mu = \lambda v_s$, with $v_s$ the vacuum
expectation value (vev) of the Singlet, that arises thus from the dynamical EWSB.\\
The NMSSM has been widely analysed in the literature (for comprehensive reviews and for lists of references,
see \cite{NMSSMreview}), considering
usually the case in which $\lambda$ is relatively small ($\lambda \lesssim 0.7$, in order to be compatible with semiperturbativity
up to the GUT scale). In this case the model has some very interesting features: first of all, the lightest scalar
does not need to satisfy the LEP bound, as long as its Singlet component is sizeable (in fact its mass can be in the
$(80\div 110)$ GeV range). This can cause the di-photon rate of the lightest Higgs particle to be a factor of $6$ higher than
the SM one due to the reduced couplings to the bottom quark, depending on the various parameters
\cite{Ellwanger:2010nf}-\cite{Cao:2011pg}. Nonetheless, it seems that in most of the parameter space the most important
decay mode of the lightest scalar particle is into a pair of light psudoscalars, so that the possibility of probing the
NMSSM
parameter space relies mainly on the ability to analyse the decays of these light pseudoscalars. Several dedicated studies
can be found in the literature (among the others, \cite{Kaplan:2011vf}-\cite{Almarashi:2010jm}-\cite{Mahmoudi:2010xp}) in
which it has been shown that with an high integrated luminosity (probably of the order of $100 fb^{-1}$) this decay
mode can be properly analysed. Very recently \cite{Domingo:2011rn}, it has been pointed out that a $A_1-\eta_b$ mixing
(where $A_1$ is the lightest pseudoscalar) can represent an efficient way to suppress the $s_1 \rightarrow A_1 A_1$
decay (with $s_1$ the lightest scalar) given an $A_1$ mass in the range $(9\div 10.5)$ GeV.\\
In contrast, in the early works about $\lambda$SUSY \cite{lambdaSUSY}, in addition to the increased Higgs boson mass
(up to about $200$ GeV), a decoupled singlet allows to have a heavier scalar with the usual couplings
of the MSSM Higgs boson: this was possible at all by suitably choosing the mass term $M_S S^2$ in the superpotential.
Moreover, due to the largish $\lambda$, the effects of the virtual Higgs bosons exchange on the
$T$ parameter are positive and automatically of the right size to compensate for the growth of both $T$ and $S$ due to
the heavier mass, so that there is agreement with the Electroweak Precision Tests (EWPT) for values of $\tan\beta$ not too
far from unity.\\
In this work, based on a scale invariant superpotential, we consider a framework which is somehow intermediate. On
the one hand we satisfy the LEP bound raising the
Higgs boson mass up to values around
$200$ GeV already at tree level using a large $\lambda$; on the other hand,
since it is no longer true in general that the singlet can be decoupled, the lightest scalar will not
simply be an
heavier MSSM Higgs boson but will have some distinctive features of the NMSSM scalars.

\section{Potential and allowed region}\label{sec:potential_allowed_region}
As already stated, the model we want to consider has the same structure as the NMSSM (for a review, see \cite{NMSSMreview}).
In order to fix the notation, we now briefly describe its structure.\\
The scale invariant model is defined by the following superpotential (that does not contain any dimensionful parameter, in
contrast to the superpotential of the MSSM):
\begin{equation}
 W = \lambda \hat{S} \hat{H}_1 \hat{H}_2 + \frac{k}{3} \hat{S}^3
\end{equation}
where $\hat S$ is a gauge singlet and $\hat{H}_{1,2}$ the usual Higgs boson fields. The corresponding scalars are defined by
\begin{equation}
 S= v_s + \frac{s_1+i s_2}{\sqrt{2}}, ~~~~
H_{1}=
\begin{pmatrix}
 v_{1}+\frac{h_{1}+ i a_{1}}{\sqrt{2}} \cr H_1^-
\end{pmatrix}, ~~~
H_2 =
\begin{pmatrix}
 H_2^+ \cr
v_{2}+\frac{h_{2}+ i a_{2}}{\sqrt{2}}
\end{pmatrix}.
\end{equation}
where $v_{1,2}$ and $v_s$ are the vevs of the two Higgs boson and of the Singlet, respectively.\\
The scalar potential is then given by
\begin{equation}\label{eq:tot_pot}
 V = V_F + V_D+ V_{SSB}
\end{equation}
where
\begin{eqnarray}\label{eq:pot_parts}
 V_F &=& |\lambda S H_2|^2 + |\lambda S H_1|^2 + |\lambda H_1 H_2 + k S^2|^2\nonumber\\
 V_D &=& \frac{g_1^2+g_2^2}{8} \left( |H_2|^2 - |H_1|^2 \right)^2+ \frac{1}{2} g_2^2 |H_1^\dagger H_2|^2\nonumber\\
 V_{SSB} &=& m_1^2 |H_1|^2 + m_2^2 |H_2|^2 + \mu_S^2 |S|^2 - \left( \lambda A S H_1 H_2 + \frac{k}{3} G S^3 + h.c.\right)
\end{eqnarray}
For simplicity in what follows we will always assume all the parameters real.\\

The striking feature of our framework is the largish value of $\lambda$ compared to the usual NMSSM case, that allows
us to increase the Higgs boson mass already at tree level acting on the F-term of the potential. Solving the relevant
RGE for $\lambda$ and $k$ \cite{Miller:2003ay}, one finds that, insisting in keeping perturbativity up to at least $10$ TeV,
at the low energy scale we must have \cite{lambdaSUSY}
\begin{equation}
 |\lambda| \lesssim 2, ~~~|k|\lesssim 0.8.
\end{equation}
in contrast to $|\lambda| \lesssim 0.7$ following from the requirement of perturbativity up to the GUT scale.
We will discuss later the sign of the relevant parameters.\\
Since the tree level upper bound on the Higgs mass is now \cite{lambdaSUSY}-\cite{Barbieri:2010pd} (see Sec. \ref{sec:mass}
for a brief discussion of the mass matrices)
\begin{equation}
 m_h^2 \leq m_Z^2 \left( \cos^2 2\beta + \frac{2 \lambda^2}{g_1^2+g_2^2} \sin^2 2\beta\right)
\end{equation}
it is clear that for high values of the $\lambda$ coupling the Higgs boson mass can easily be raised at the level of
$(200\div 250)$ GeV, so that from now on we will use $\lambda=2$.\\

We will now discuss all the relevant constraints that we have to impose in order to find the allowed region in parameter
space, namely the stability of the potential, conservation of electromagnetism, absence of spontaneous CP violation and
absence of unrealistic minima.
\subsection{Stability, conservation of Electromagnetism and absence of spontaneous CP violation}\label{sec:stability_etc}
As usual in the NMSSM the potential is always stable, while the condition of unbroken electromagnetism is
\begin{equation}
 \lambda^2 |v_s|^2 + \frac{g_2^2}{4}\left( |v_1|^2+|v_2|^2\right) + \frac{g_1^2}{4}\left( |v_1|^2-|v_2|^2\right)
+m_1^2>0
\end{equation}
The issue of spontaneous CP violation has to be treated carefully, since the requirement of real parameters
does not imply
CP conservation in the scalar sector. This follows from the fact that while with an
$\mathrm{SU(2)}\times \mathrm{U(1)}$ transformation we can always set $v_2>0$, in general both $v_1$ and $v_s$ are
complex parameters ($v_1= |v_1| e^{i \varphi_1}$, $v_s = |v_s| e^{i \varphi_s}$). We thus want
$(\varphi_1, \varphi_s)=(0,0)$ to be an absolute minimum. The only part of the neutral scalar potential that
depends on the two phases is the one trilinear in the scalar fields:
\begin{equation}\label{eq:CP_viol_pot}
 V_{0,min} = - 2 \left( \lambda A v_1 v_2 v_s \cos(\varphi_1+\varphi_s)+\frac{k}{3} G v_s^3 \cos(3 \varphi_s) -\lambda k v_1 v_2 v_s^2 \cos(2\varphi_s -\varphi_1)\right)
\end{equation}
which follows from eqs. (\ref{eq:tot_pot})-(\ref{eq:pot_parts}) after the insertion of the complex vevs.\\
From the matrix of second derivatives one obtains the conditions for $(\varphi_1, \varphi_s)=(0,0)$ to be a local minimum:
\begin{eqnarray}\label{eq:CP}
 0 &<& 2 v_s \left( 3 k G v_s^2 + \lambda v_1 v_2 (2 A - 5 k v_s)\right)\nonumber\\
 0 &<& 12 \lambda k v_1 v_2 v_s^3 \left( A G v_s - 3 \lambda A v_1 v_2 - k G v_s^2\right)
\end{eqnarray}
Although it can be shown that non trivial extremal points exists, these points are always local maxima
\cite{Romao:1986jy}. Depending on the signs of the parameters involved, from eq. (\ref{eq:CP}) we can summarize the situation as follows
\cite{Cerdeno:2004xw}:
\begin{itemize}
\item if $k<0$ and $\mathrm{sgn}(v_s) = \mathrm{sgn}(A) = -\mathrm{sgn}(G)$, then $(\varphi_1, \varphi_s)=(0,0)$ is always a minimum;
\item if $k<0$ and $\mathrm{sgn}(v_s) = -\mathrm{sgn}(A) = -\mathrm{sgn}(G)$ then $(\varphi_1, \varphi_s)=(0,0)$ is a minimum if
$|G| > 3\lambda v_1 v_2 |A|/\left(-|v_s A|+|k| v_s^2\right)$ and the denominator is positive;
\item if $k<0$ and $\mathrm{sgn}(v_s) = \mathrm{sgn}(A) = \mathrm{sgn}(G)$ then $(\varphi_1, \varphi_s)=(0,0)$ is a minimum if
$|G| < 3\lambda v_1 v_2 |A|/\left(|v_s A|+|k| v_s^2\right)$;
\item if $k>0$ and $\mathrm{sgn}(v_s) = \mathrm{sgn}(A) = \mathrm{sgn}(G)$ then $(\varphi_1, \varphi_s)=(0,0)$ is a minimum if
 $|G| > 3\lambda v_1 v_2 |A|/\left(|v_s A|-|k| v_s^2\right)$ and a positive denominator;
\end{itemize}
Comparing now the point $(\varphi_1, \varphi_s)=(0,0)$ with $(\varphi_1, \varphi_s)=(0, \pm \pi,0, \pm \pi)$, it is easy
to show that in the cases previously listed $(\varphi_1, \varphi_s)=(0,0)$ is always an absolute minimum and not only
a local minimum.\\

It is useful to analyse if all the signs of the parameters are physical or
if we can fix some of them. To this end, we notice that the neutral part of the potential is invariant under the
transformations
\begin{eqnarray}
 (\lambda, v_1) &\rightarrow & -(\lambda, v_1)\nonumber\\
 (\lambda, k, v_s) &\rightarrow & -(\lambda, k, v_s).
\end{eqnarray}
This allow us to fix the sign of $\lambda$ and $v_1$, but not those of $A$, $G$, $v_s$ and $k$ that are thus physical.
In what follows we will always choose $\lambda, ~v_1 >0$ (and $v_2>0$ as already said).\\

\begin{figure}[Htb]
 \begin{center}
  \begin{tabular}{cc}
   \includegraphics[width=.45\textwidth]{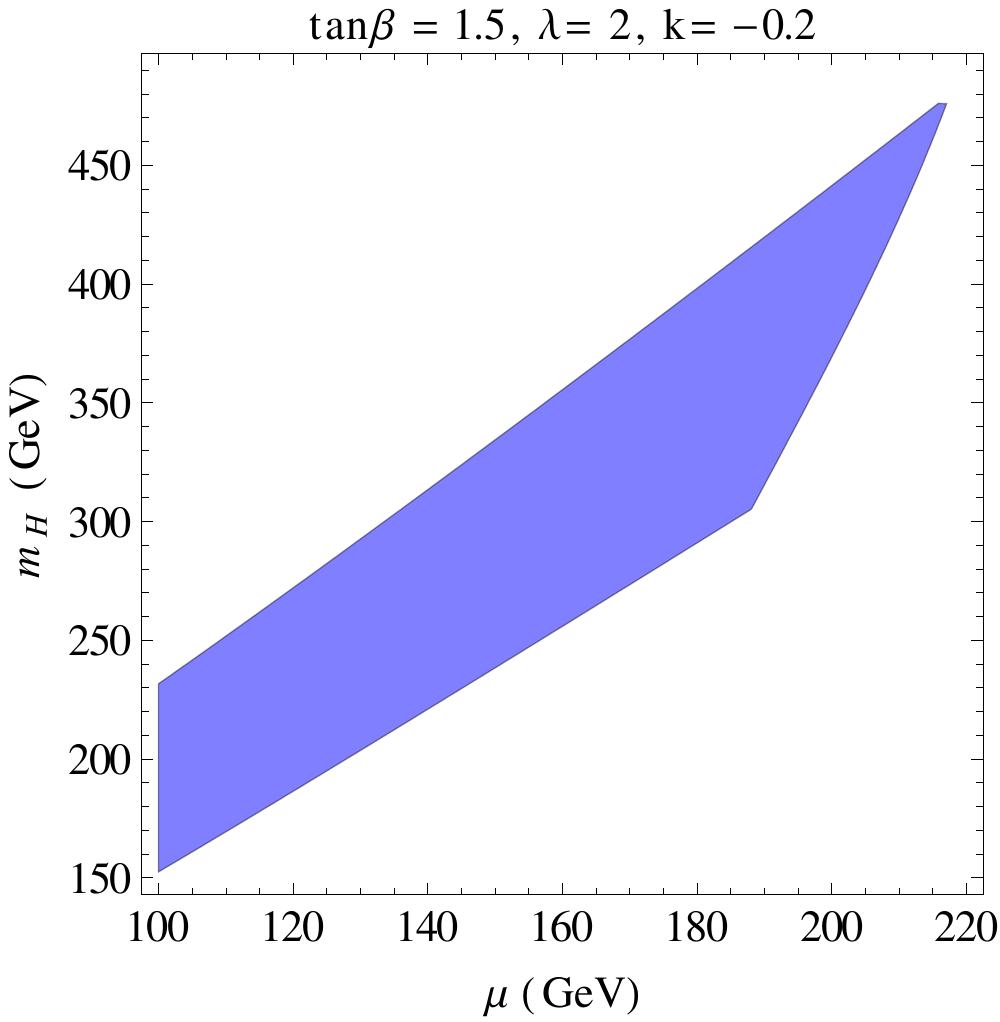} &
   \includegraphics[width=.45\textwidth]{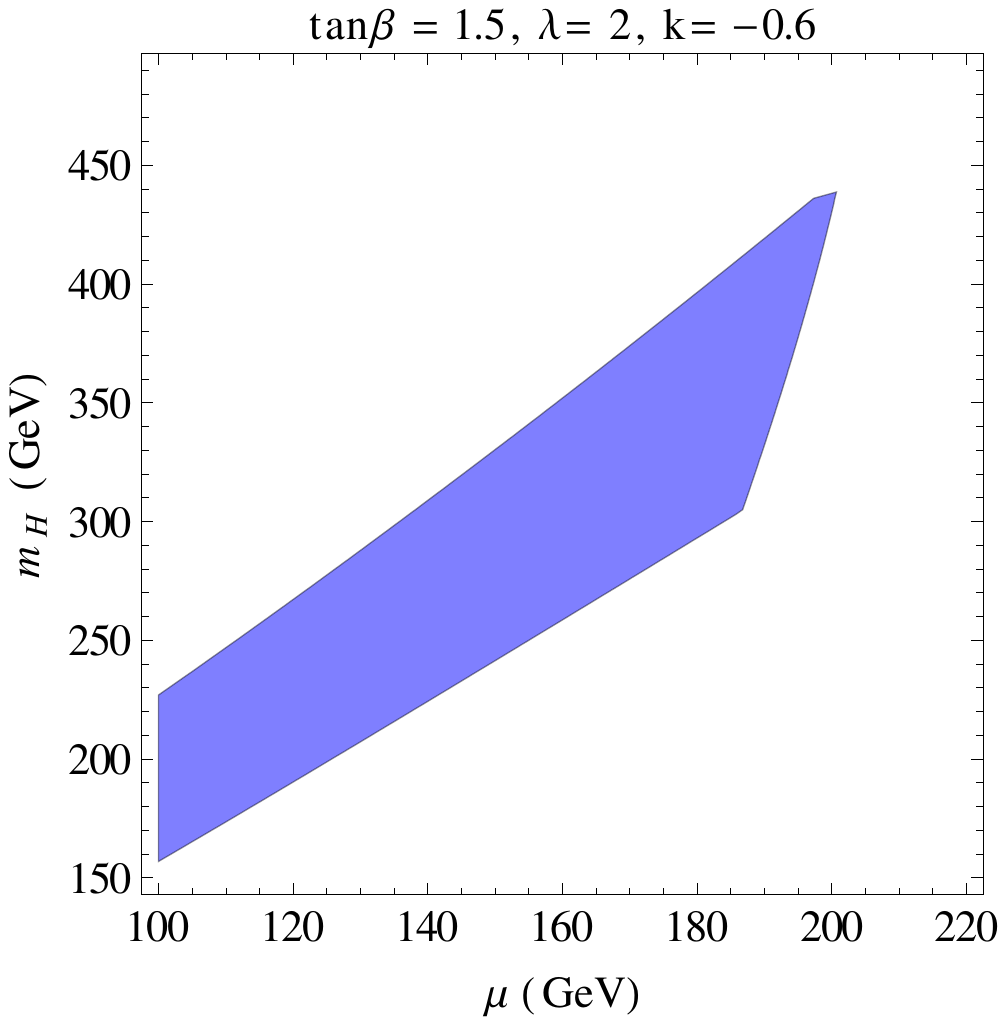}
  \end{tabular}
 \end{center}
\caption{Coloured region: allowed region in parameter space in the $k<0$ case once the constraints coming from the requirement
of no spontaneous CP violation and from the absence of unrealistic minima are taken into
 account. In the $k>0$ case only a very small region (or no region at all) survives. The allowed region is symmetric for
$\mu \rightarrow -\mu$.}
\label{fig:reg}
\end{figure}

\subsection{Unrealistic minima}
The last potentially dangerous issue we have to care about is the presence of unrealistic minima of the potential that
we don't want to be the absolute minimum. This minima are particularly dangerous in the case of large $\lambda$
\cite{Kanehata:2011ei}, while they are not a particular problem in the usual NMSSM with low values of $\lambda$.\\
We notice that in general, barring special relationships between the parameters, when two of the scalar fields acquire their
vevs also the third one must develop a vev to satisfy the minimum conditions.  This leaves us with the analysis of cases
in which either one or three scalar fields have non vanishing vev.\\
In the case $v_1,~v_2,~v_s\neq 0$, the minimization gives ($\tan\beta=v_2/v_1$)
\begin{eqnarray}\label{eq:real_min}
 \tan^2\beta &=& \frac{2\left( m_1^2+ \lambda^2 v_s^2\right) + m_Z^2}{2\left( m_2^2+ \lambda^2 v_s^2\right) + m_Z^2}\nonumber\\
 \lambda^2 v^2 &=& \frac{\lambda \left(A-k v_s\right)}{\sin\beta \cos\beta}- \left(m_1^2+m_2^2\right) - 2 \lambda^2 v_s^2\nonumber\\
 0 &=& 4 k^2 v_s^3 - 2 G k v_s^2 + 2 v_s\left( \lambda v^2 (k \sin2\beta +\lambda)+\mu_S^2\right) - A v^2 \lambda \sin2\beta
\end{eqnarray}
from which we get that the minimum value of the potential is given by
\begin{equation}\label{eq:true_min}
 V_{min}^{true} = \mu_S^2 v_s^2 + k^2 v_s^4- \frac{2}{3} k G v_s^3 -\lambda^2 v^4 \sin^2\beta \cos^2\beta -
\frac{m_Z^2 v^2}{4} \cos{2\beta}
\end{equation}
with $\mu_S^2$ given by the last eq. in (\ref{eq:real_min}) as a function of $v_s$.\\
Regarding the minima with two of the three vevs equal to zero, we first of all consider the case $v_1 =0 = v_2$. It is
straightforward to find that the minimum of the potential is given by
\begin{equation}\label{eq:VS}
 V_S = -\frac{1}{6} v_s^2 \left( k G v_s - 3 \mu_S^2\right), ~~~~~v_s = \frac{k G \pm \sqrt{k^2 (G^2 - 8 \mu_S^2)}}{4 k^2}
\end{equation}
The other cases we have to consider are those in which $v_{1,2}=0=v_s$. Taking first $v_1=0$, we find that the minimum of the
potential is given by
\begin{equation}\label{eq:VH2}
 V_{H2} = m_2^2 v_2^2+  \frac{g_1^2+g_2^2}{8} v_2^4, ~~~~~v_2^2 = - 4 \frac{m_2^2}{g_1^2+g_2^2}
\end{equation}
An analogous expression can be found in the $v_2=0$ case, for which the relevant formulae can be obtained setting
$m_2\rightarrow m_1$. We will call $V_{H1}$ the value of the potential in this minimum.\\
Since we want eq. (\ref{eq:true_min}) to be the true minimum, we impose the following conditions:
\begin{equation}\label{eq:minima_cond}
 V_{S, H1, H2} > V_{min}^{true}, ~~~0 >V_{min}^{true}.
\end{equation}
where the last inequality follows considering the trivial minimum $(v_1, v_2, v_s)=(0,0,0)$.\\

\subsection{Allowed region in parameter space}
\begin{figure}[h!tb]
 \begin{center}
  \begin{tabular}{cc}
   \includegraphics[width=.35\textwidth]{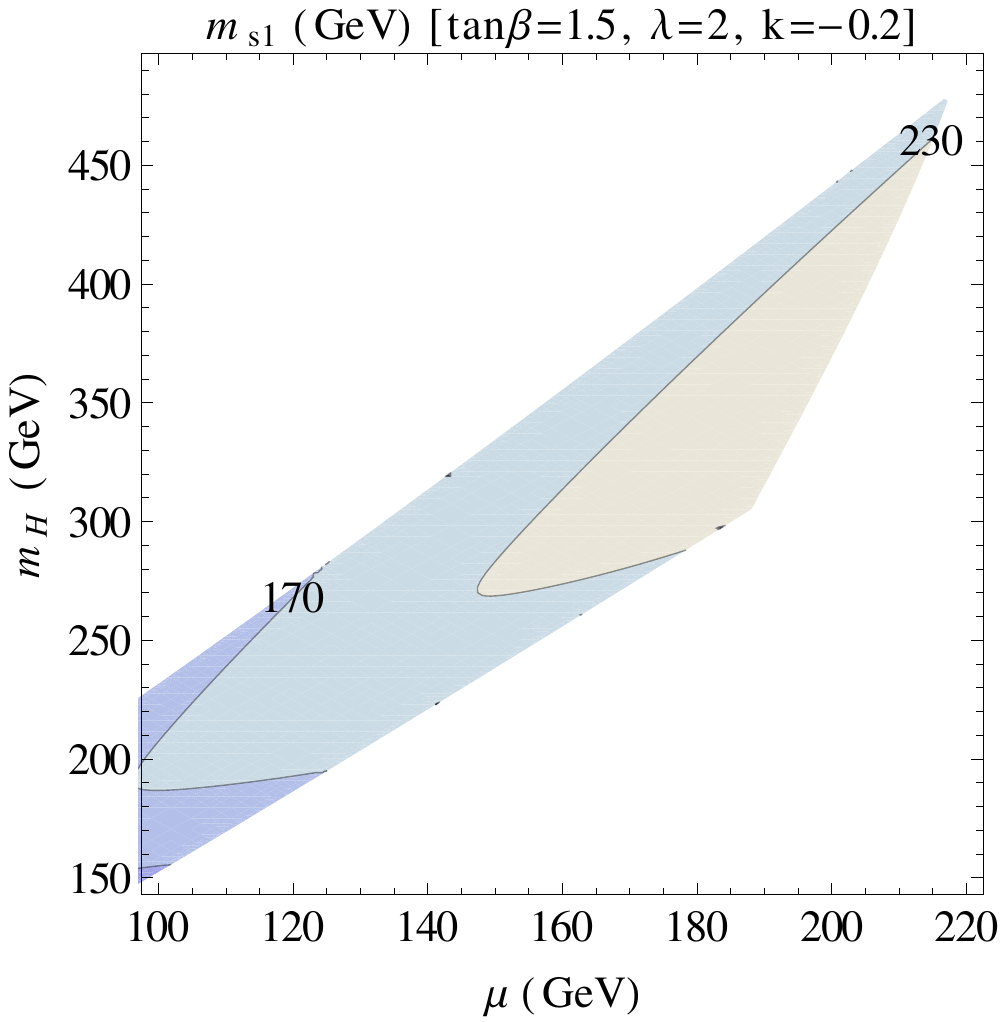} &
   \includegraphics[width=.35\textwidth]{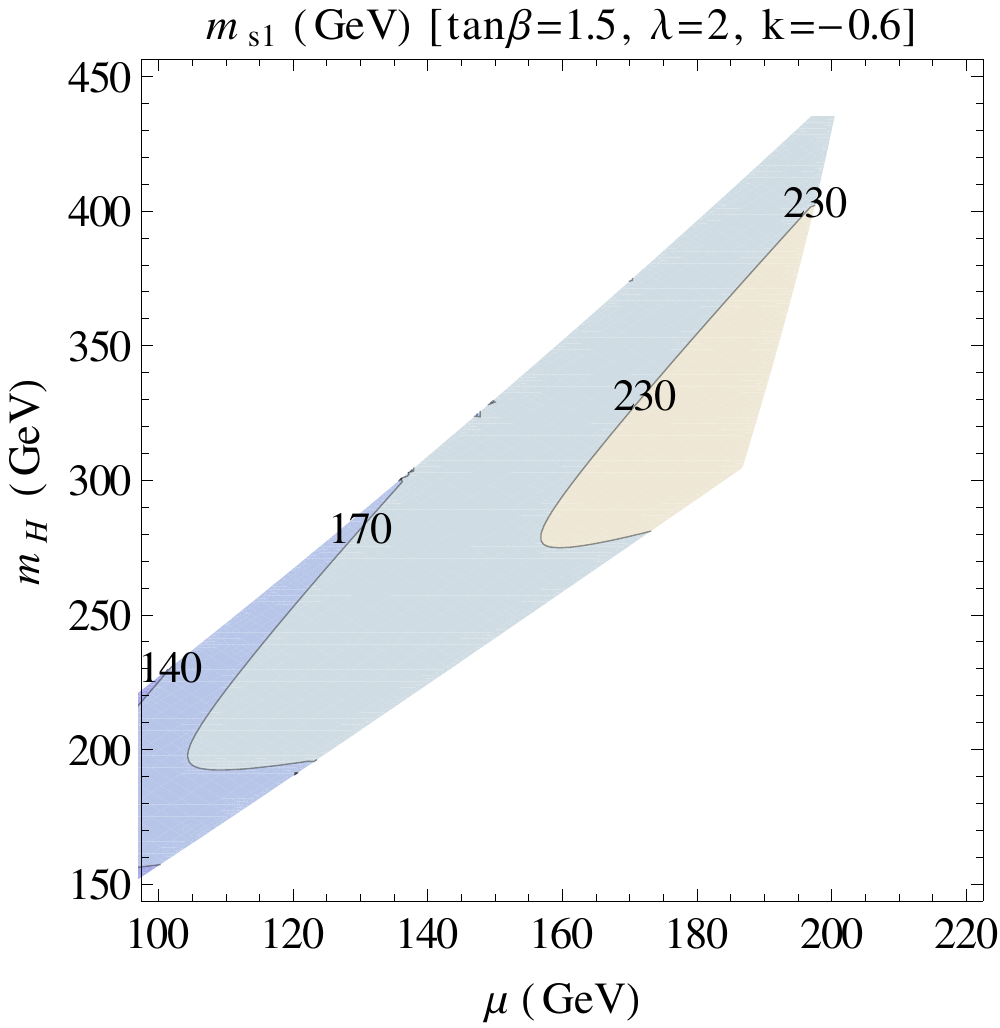}\\
   \includegraphics[width=.35\textwidth]{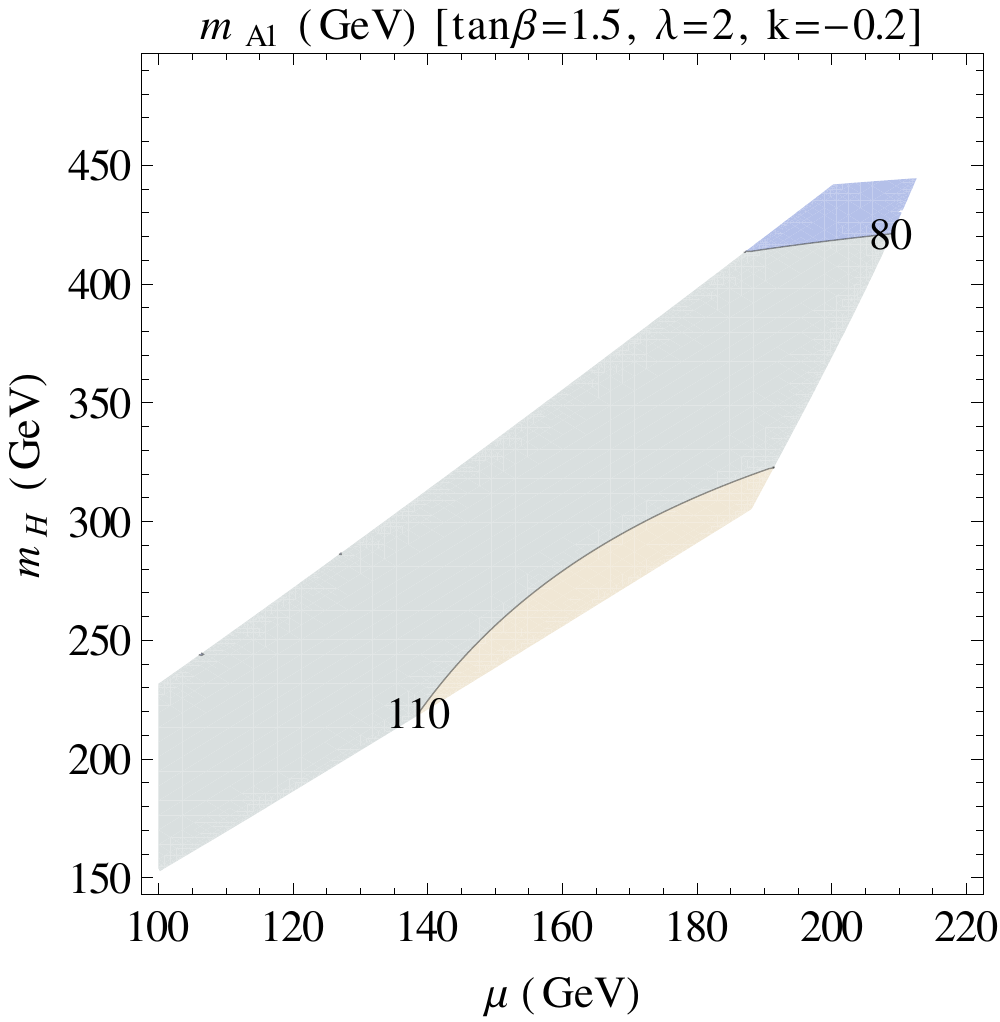} &
   \includegraphics[width=.35\textwidth]{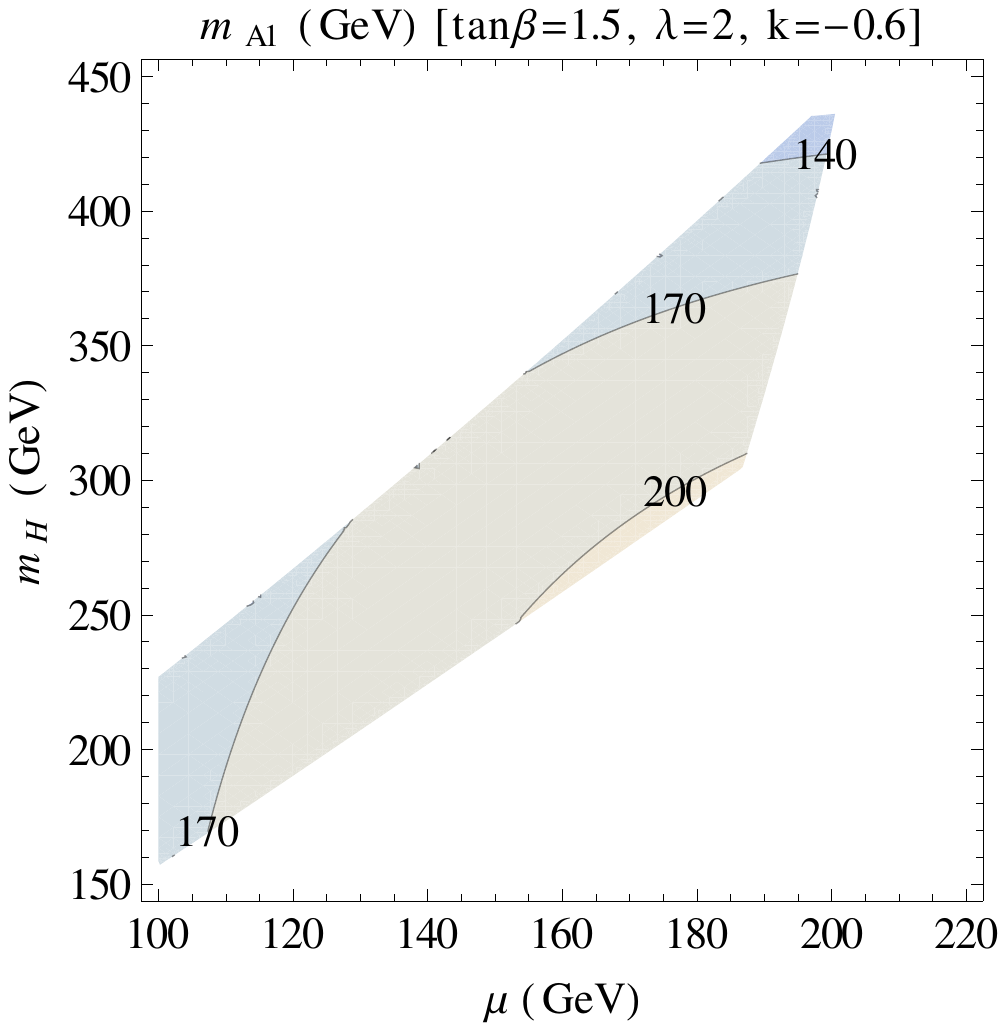}\\
   \includegraphics[width=.35\textwidth]{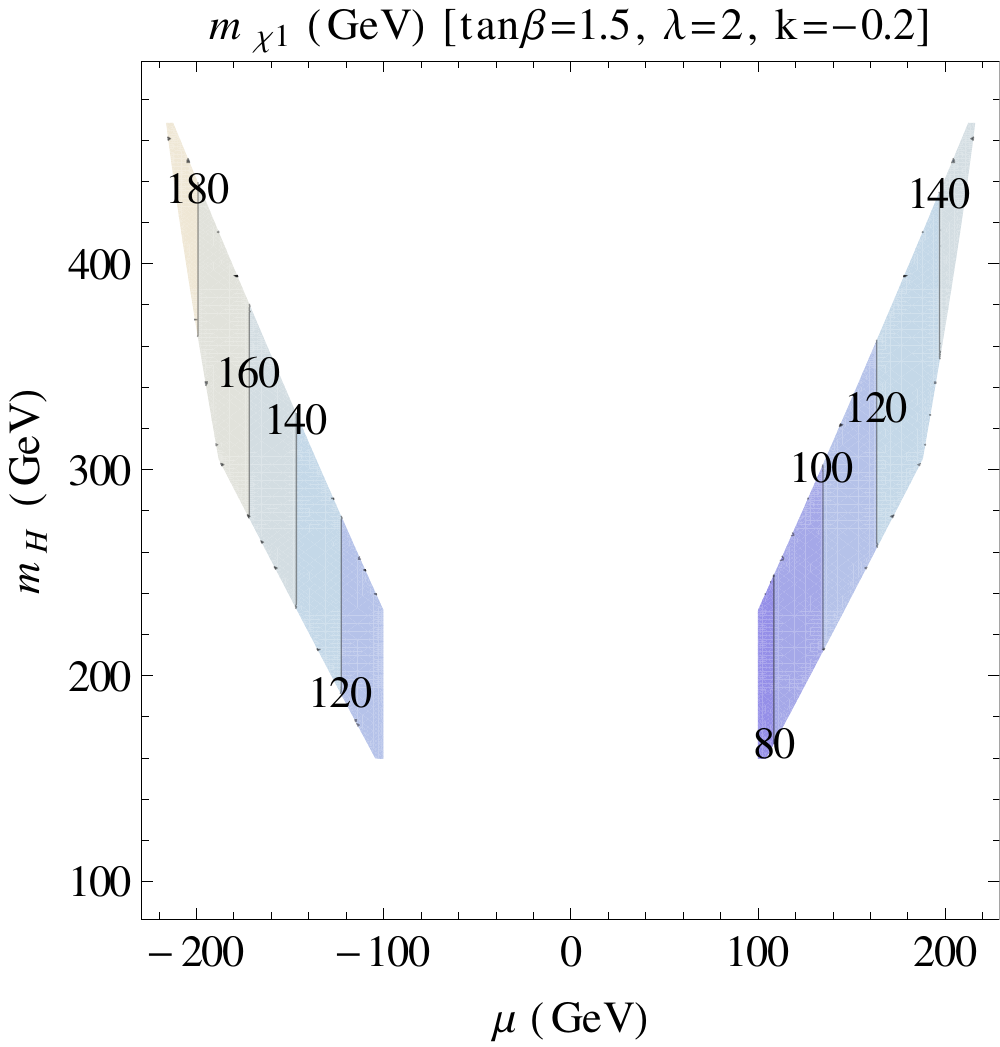} &
   \includegraphics[width=.35\textwidth]{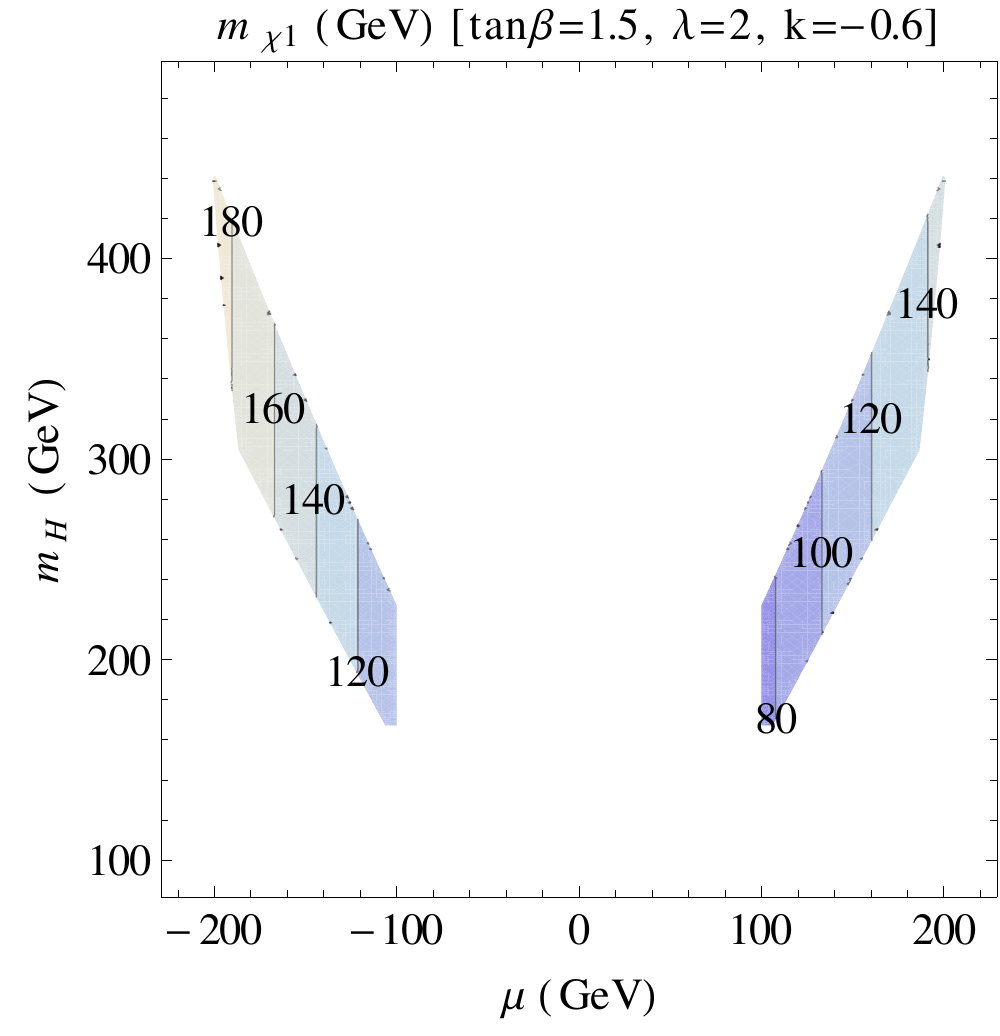}\\
  \end{tabular}
 \end{center}
\caption{Mass of the lightest scalar (upper panels), pseudoscalar (middle panels) and neutralino (lower panels). The gaugino
mass parameters are fixed as
$M_1=200$ GeV
and $M_2=2$ TeV, so that $\mu$ coincides with the chargino mass.}
\label{fig:masses}
\end{figure}
Before discussing the allowed region obtained imposing the different constraints discussed up to now,
a comment about our choice of the free
parameters is in order. Since $v\simeq 174$ GeV is fixed, we can use eq. (\ref{eq:real_min}) to eliminate $m_{1,2}$
and $\mu_S$ in favour of $\tan\beta$ and $\mu \equiv \lambda v_s$; at the same time, the charged
Higgs boson mass is given by \cite{Masses}
\begin{equation}\label{eq:mHcharged}
 m_H^2 = m_W^2 -\lambda^2 v^2 +\frac{\mu \left(A - \frac{k}{\lambda}\mu\right)}{\sin\beta \cos\beta}
\end{equation}
so that we can use the previous expression to re-express $A$ in terms of $m_H$\footnote{Notice that the condition of
unbroken electromagnetism discussed at the beginning of Sec. \ref{sec:stability_etc} can be written in the new variables
simply as $m_H>0$.}. Taking for simplicity $A=G$ (we will
comment later on what happens relaxing this condition), we are left
with five parameters: $\lambda$, $k$, $\tan\beta$, $\mu$ and $m_H$.\\

The allowed region can be obtained combining eq. (\ref{eq:CP}) (absence of spontaneous CP
violation) and eq. (\ref{eq:minima_cond}) (absence of unrealistic minima). The $k>0$ case is almost completely ruled out,
since only a very small portion of the parameter space (or no region at all, depending on the value of $k$) is allowed; in
what follows we will thus neglect this case. On the contrary, in the $k<0$ case there is an allowed region, which is
shown in Fig. \ref{fig:reg} for two different values of $k$. We stress that the plot is symmetric under
$\mu\rightarrow -\mu$
so that also the symmetric region for $\mu<0$ is allowed. We take as lower bound $\mu \gtrsim 100$ GeV having in mind
the situation in which $M_2 \gg \mu$, so that
$\mu$ is the mass of the lightest chargino and the bound $m_{\chi^+} > 94$ GeV applies \cite{Nakamura:2010zzi}. Regarding
the charged Higgs boson mass, in order to satisfy the $b\rightarrow s\gamma$ constraint, a lower bound $m_H \gtrsim 350$ GeV
is usually quoted \cite{Gambino:2001ew}. However, this bound does not keep into account possible destructive contributions,
coming \emph{i.e.} from the stop-chargino loop, that can allow lower charged Higgs mass \cite{Mahmoudi:2010xp}. For this
reason, we keep the PDG lower bound $m_H\gtrsim 79.3$ GeV at $95\%$ C.L. \cite{Nakamura:2010zzi}.
\begin{figure}[h!t]
 \begin{center}
  \begin{tabular}{cc}
   \includegraphics[width=.35\textwidth]{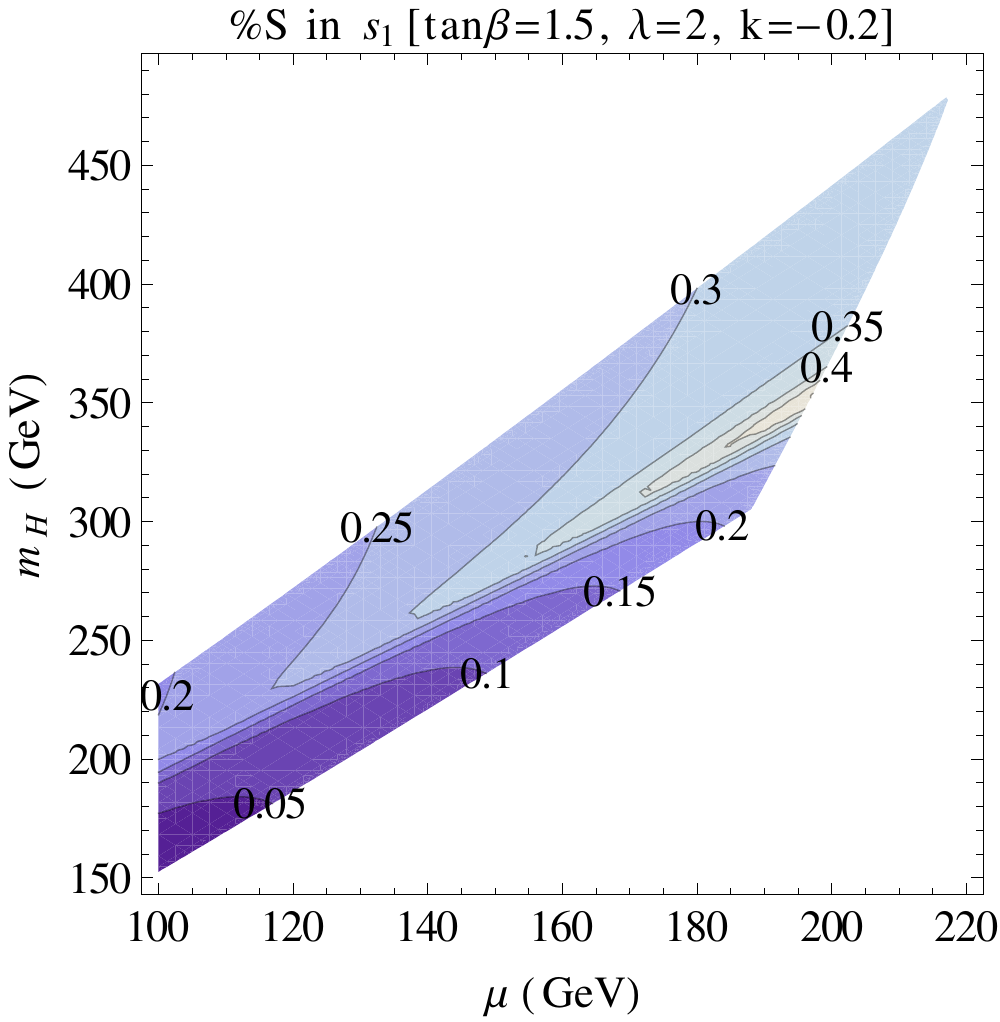} &
   \includegraphics[width=.35\textwidth]{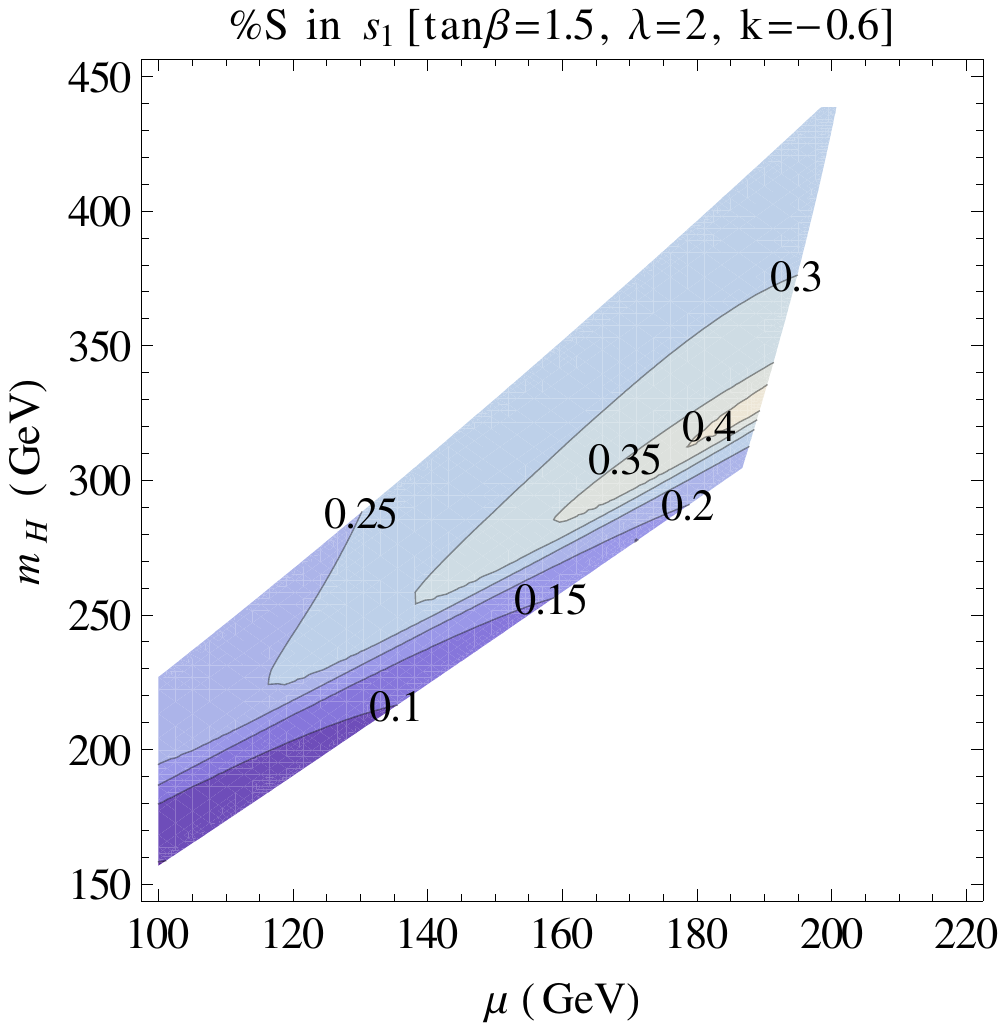}\\
   \includegraphics[width=.35\textwidth]{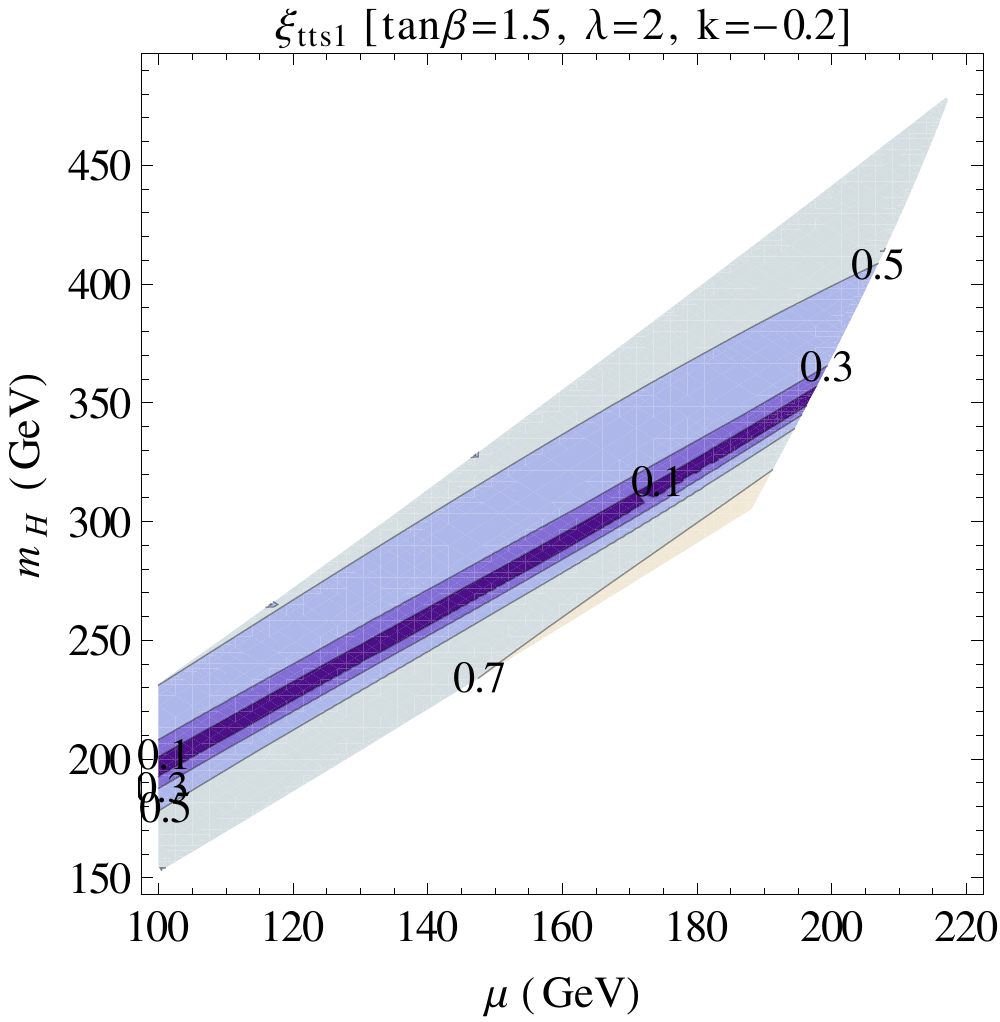} &
   \includegraphics[width=.35\textwidth]{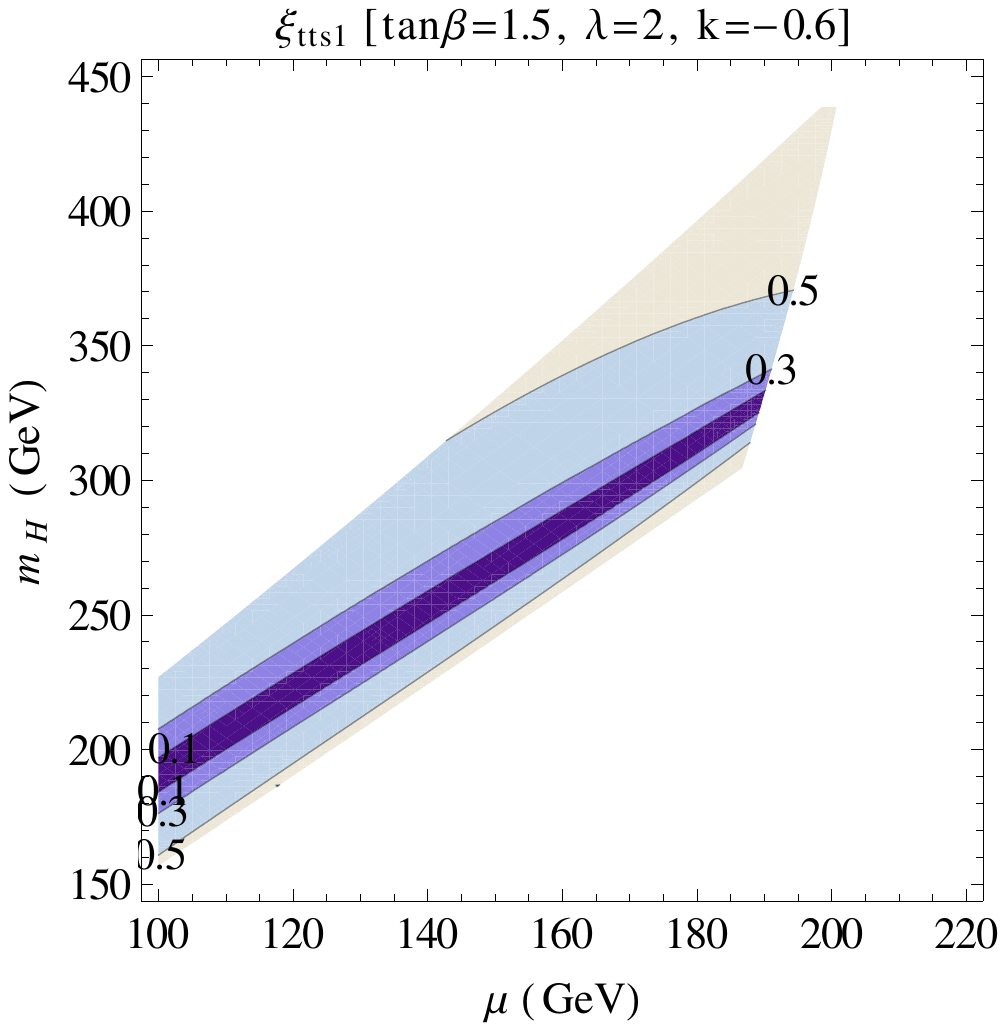}\\
   \includegraphics[width=.35\textwidth]{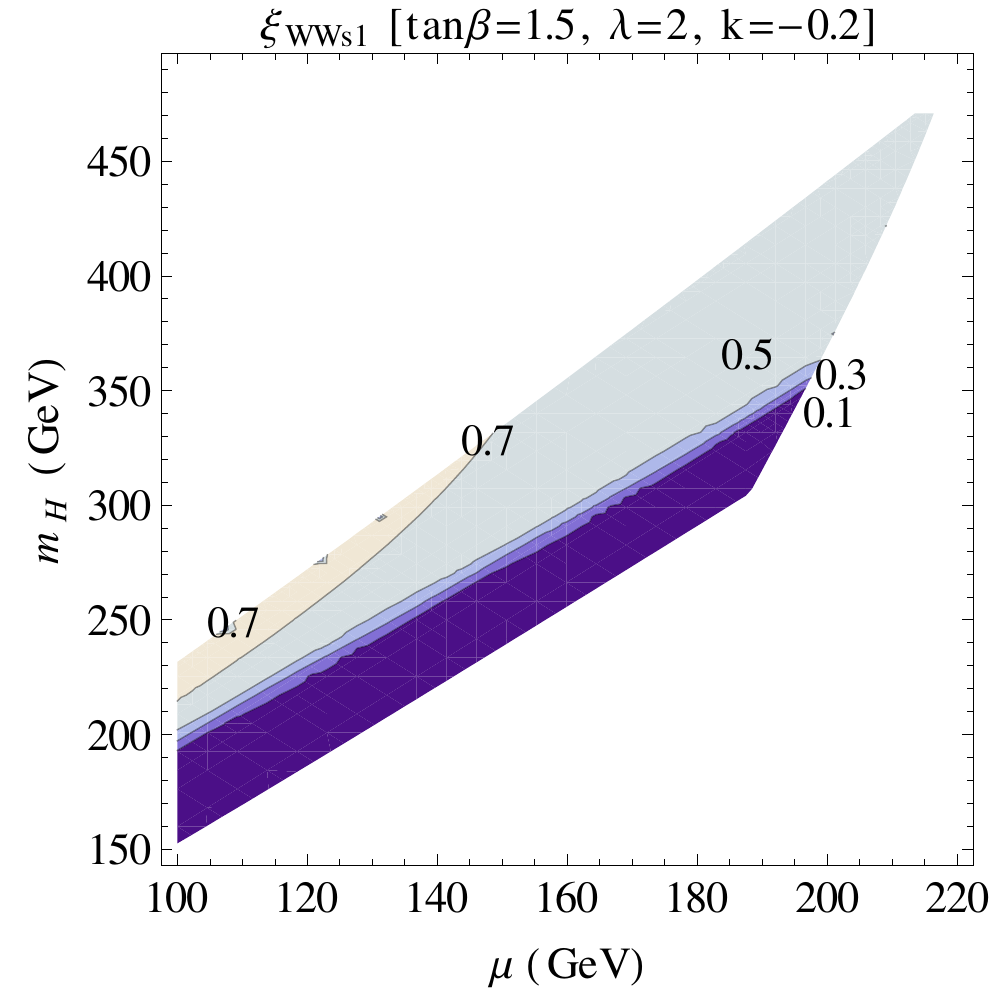} &
   \includegraphics[width=.35\textwidth]{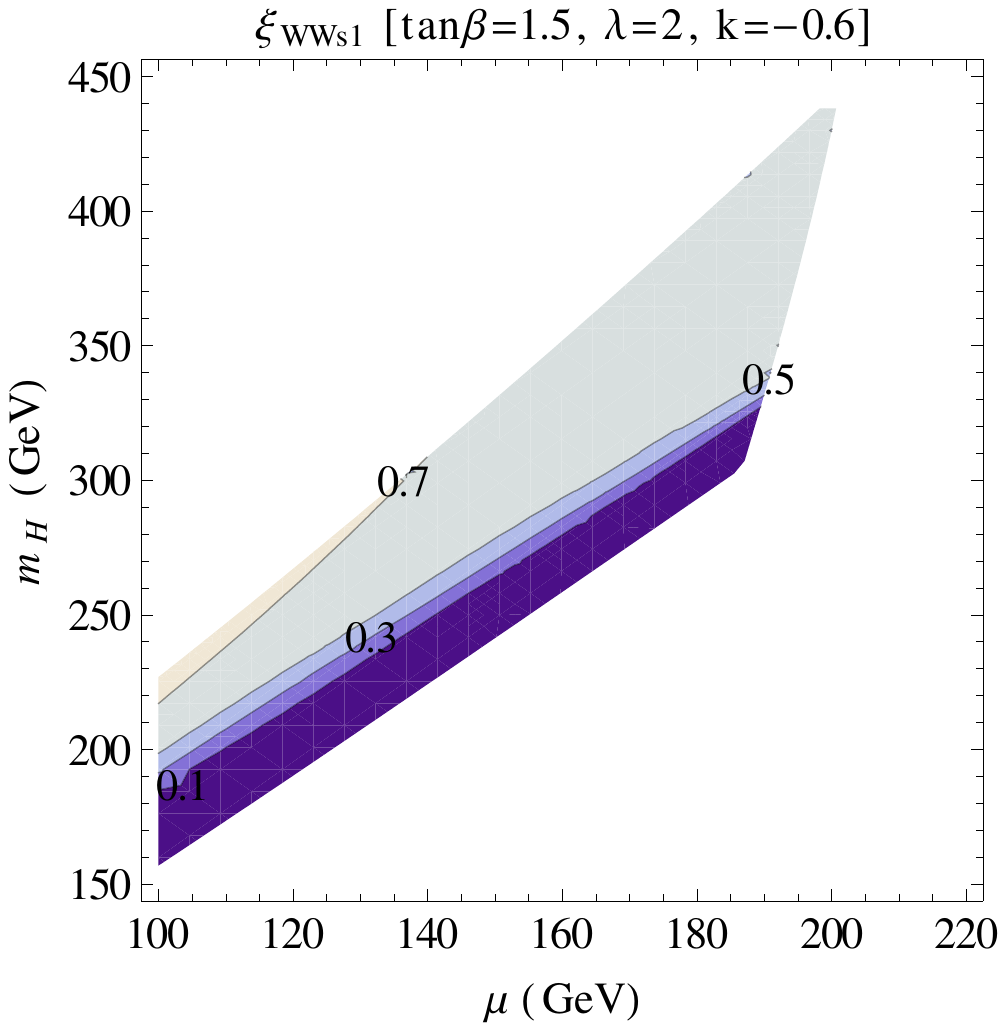}\\
  \end{tabular}
 \end{center}
\caption{Singlet component in the lightest scalar (upper panel) and ratio of cross sections with respect to the SM one
in the case of Gluon-Gluon fusion (middle panel) and Vector Boson Fusion (lower panel), see eq. \ref{eq:xitts1}.}
\label{fig:percS}
\end{figure}
\section{Mass spectrum}\label{sec:mass}
The physical scalar sector of our theory consists of three scalars $s_i$, $i=1, \dots, 3$, two pseudoscalars $A_i$,
$i=1,2$, (both with masses labeled in increasing order), and by one charged scalar $H^+$.\\
As already said, minimizing the potential one finds that the mass of the charged Higgs boson is given by eq.
(\ref{eq:mHcharged}), while the mass matrices in the scalar and pseudoscalar sector are given respectively by (in the
$(h_1, h_2, s_1)$ and $( \sin\beta a_1 + \cos\beta a_2, s_2)$ basis) \cite{Masses}
{\small
\begin{equation}
 M_S^2 =
\begin{pmatrix}
 c_\beta^2 m_Z^2 + s_\beta^2 m_A^2 & c_\beta s_\beta (2 v^2 \lambda^2 -m_Z^2 -m_A^2) & \mu v (2 \lambda c_\beta + s_\beta k) -
s_\beta^2 c_\beta v \lambda \frac{m_A^2}{\mu}
\cr
\cdot & s_\beta^2 m_Z^2 + c_\beta^2 m_A^2 & -\frac{v \lambda c_\beta^2 s_\beta m_A^2}{\mu}+ v\mu (2 \lambda s_\beta + k c_\beta)
\cr
\cdot & \cdot & 4 \frac{k^2}{\lambda^2}\mu^2 - \left( \frac{s_\beta c_\beta m_A^2}{\mu}+ \frac{k}{\lambda}\mu\right)
\frac{k}{\lambda}\mu + \lambda^2 v^2 s_{2\beta}^2 \frac{m_A^2}{4 \mu^2}+\frac{\lambda k}{2} v^2 s_{2\beta}
\end{pmatrix}
\end{equation}
\begin{equation}
M_P^2 =
\begin{pmatrix}
m_A^2 & \frac{\lambda v s_\beta c_\beta m_A^2}{\mu} + 3 k v \mu
\cr
\cdot & \lambda^2 v^2 s_\beta^2 c_\beta^2 \frac{m_A^2}{\mu^2} - 3 \lambda k s_\beta c_\beta v^2 + 3
\left( \frac{s_\beta c_\beta m_A^2}{\mu}+ \frac{k}{\lambda}\mu\right) \mu \frac{k}{\lambda}
\end{pmatrix}
\end{equation}}
where $m_A^2 =  m_H^2 + \lambda^2 v^2- m_W^2$ is the mass of the MSSM pseudoscalar and $c_\beta \equiv \cos\beta$, $s_\beta
\equiv \sin\beta$.\\

In the fermion sector, the higgsinos have a mixing with the singlino while the gauginos mix only with the higgsinos, so that the relevant
$5\times 5$ mass matrix in the $( \tilde{B}, \tilde{W}, \frac{\tilde{H}_1-\tilde{H}_2}{\sqrt{2}} ,
\frac{\tilde{H}_1+\tilde{H}_2}{\sqrt{2}}
, \tilde{S} )$ basis is given by
 \begin{equation}\label{eq:neutr_mass_matr}
M_\chi =
\begin{pmatrix}
 M_1 & 0 & - \frac{m_Z ( c_\beta + s_\beta) s_W}{\sqrt{2}} &  \frac{m_Z ( c_\beta - s_\beta) s_W}{\sqrt{2}} & 0
\cr
\cdot & M_2 & \frac{m_Z ( c_\beta + s_\beta) c_W}{\sqrt{2}} & \frac{m_Z ( c_\beta -  s_\beta) c_W}{\sqrt{2}} & 0
\cr
\cdot & \cdot & \mu & 0 & \frac{v}{\sqrt{2}} \lambda (c_\beta -s_\beta)
\cr
\cdot & \cdot & \cdot & -\mu & -\frac{v}{\sqrt{2}} \lambda (c_\beta + s_\beta)
\cr
\cdot & \cdot & \cdot & \cdot & - 2 { k \over \lambda} \mu
\end{pmatrix}
\end{equation}
The mass of the lightest scalar, pseudoscalar and neutralino in shown in Fig. \ref{fig:masses} for the same cases considered
in Fig. \ref{fig:reg}. For scalars and pseudoscalars only the $\mu>0$ region is shown, since the plot is symmetric under
$\mu \rightarrow -\mu$. As one can see, the value of $k$ affects only
marginally the scalars masses, while this is not true for pseudoscalars, in which case the value of $k$ matters.
On the other hand, the neutralino masses changes significantly according to the sign of $\mu$: this can be explained
considering that the lightest neutralino has a non negligible Bino component due to our choice $M_1=200$ GeV and $M_2=2$ TeV,
while in the case of decoupled gauginos it is simple to see that there would be only a small change between the $\mu>0$
and the $\mu<0$ region.\\

As last comment, let us discuss the Electroweak Precision Tests (EWPT), since naively one could expect problems
from the high Higgs boson mass. We have checked that, as long as we keep $1.5 \lesssim \tan\beta \lesssim 2$ both the
higgsino and the top-stop contributions do not give too high contributions to the $T$ and $S$ parameters \cite{lambdaSUSY}.
\section{Scale invariant $\lambda$SUSY at the LHC}\label{sec:phenomenology}
\begin{figure}[H!t]
 \begin{center}
  \begin{tabular}{cc}
   \includegraphics[width=.36\textwidth]{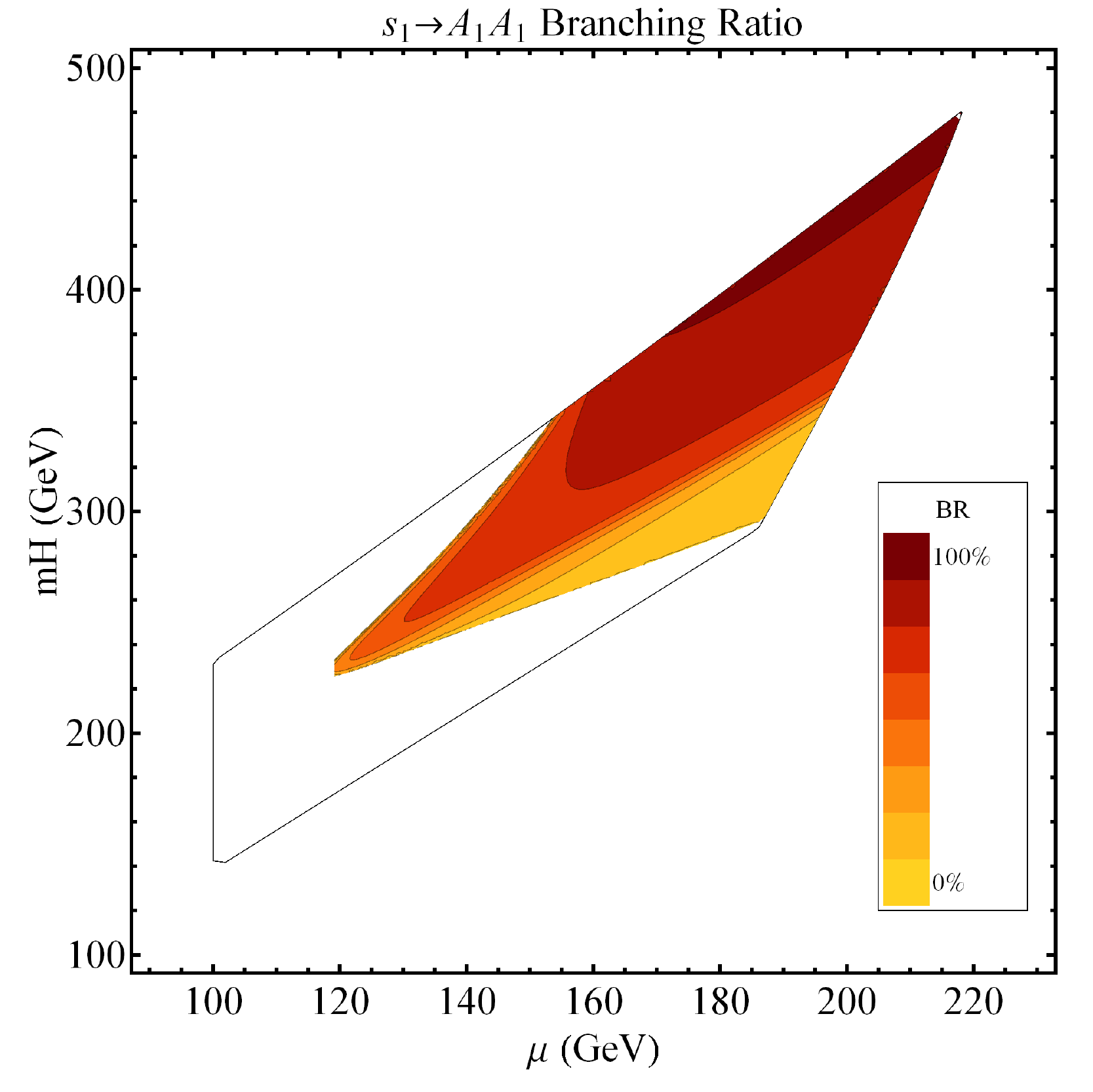} &
   \includegraphics[width=.36\textwidth]{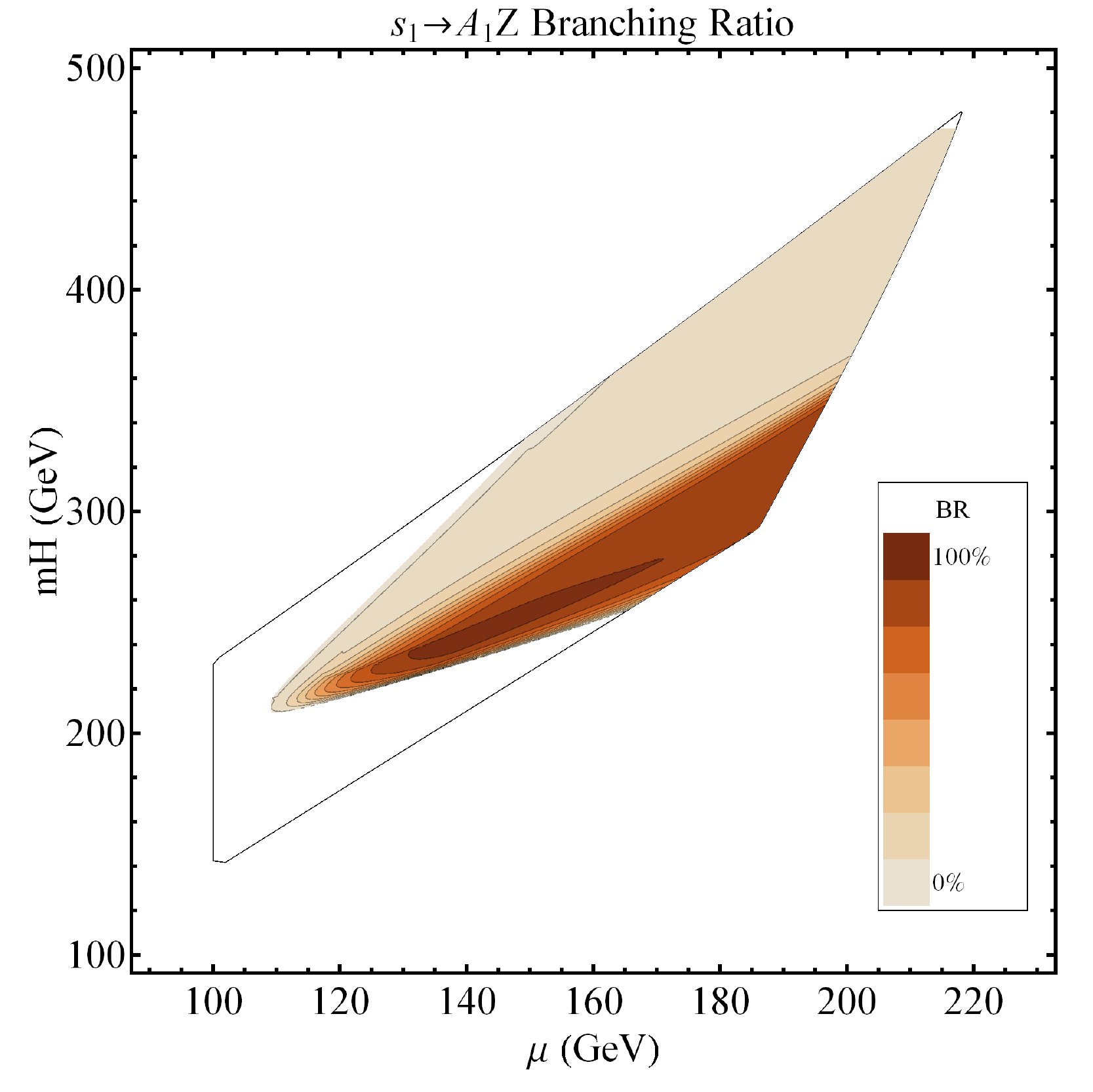}\\
   \includegraphics[width=.36\textwidth]{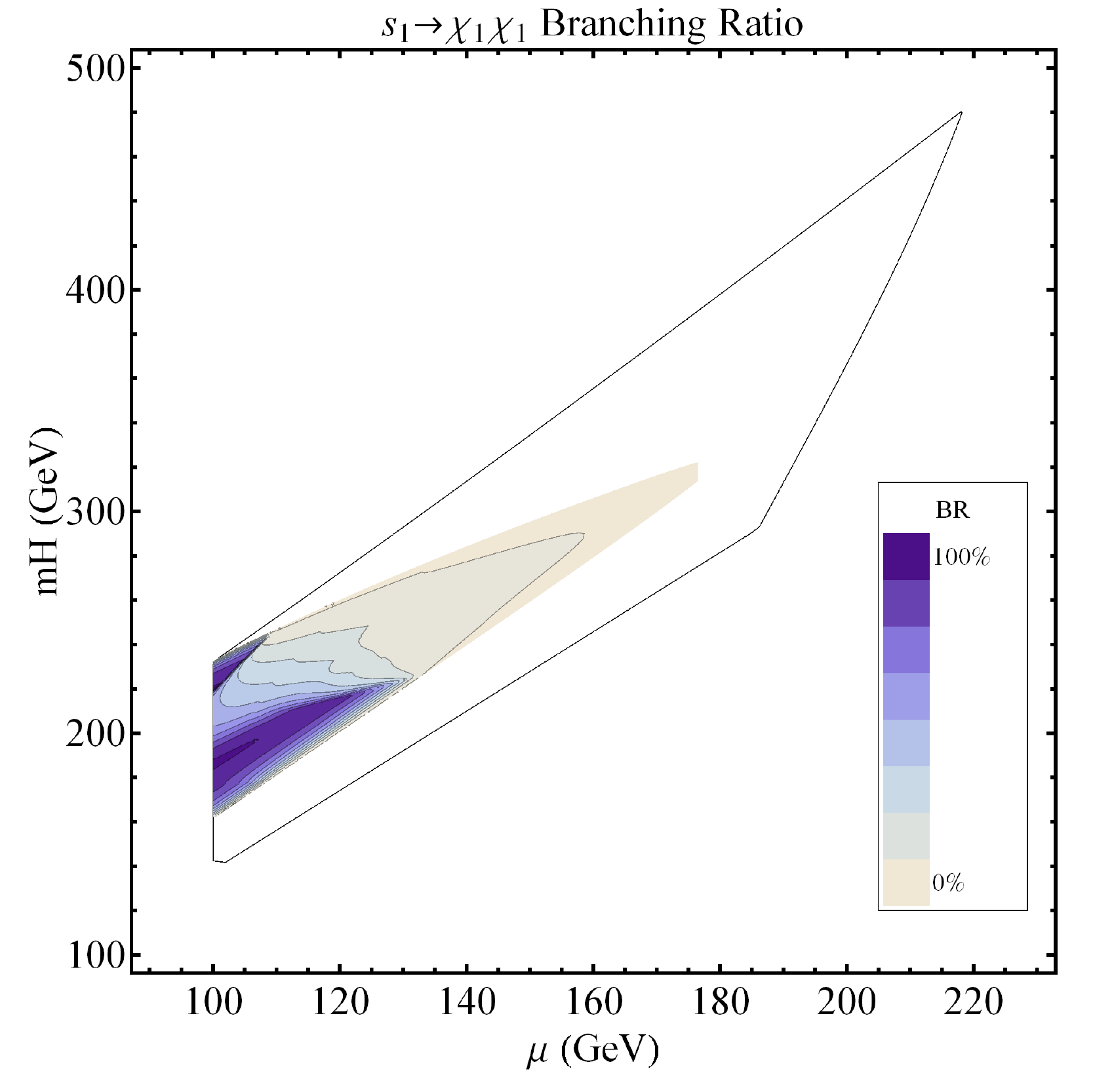} &
   \includegraphics[width=.36\textwidth]{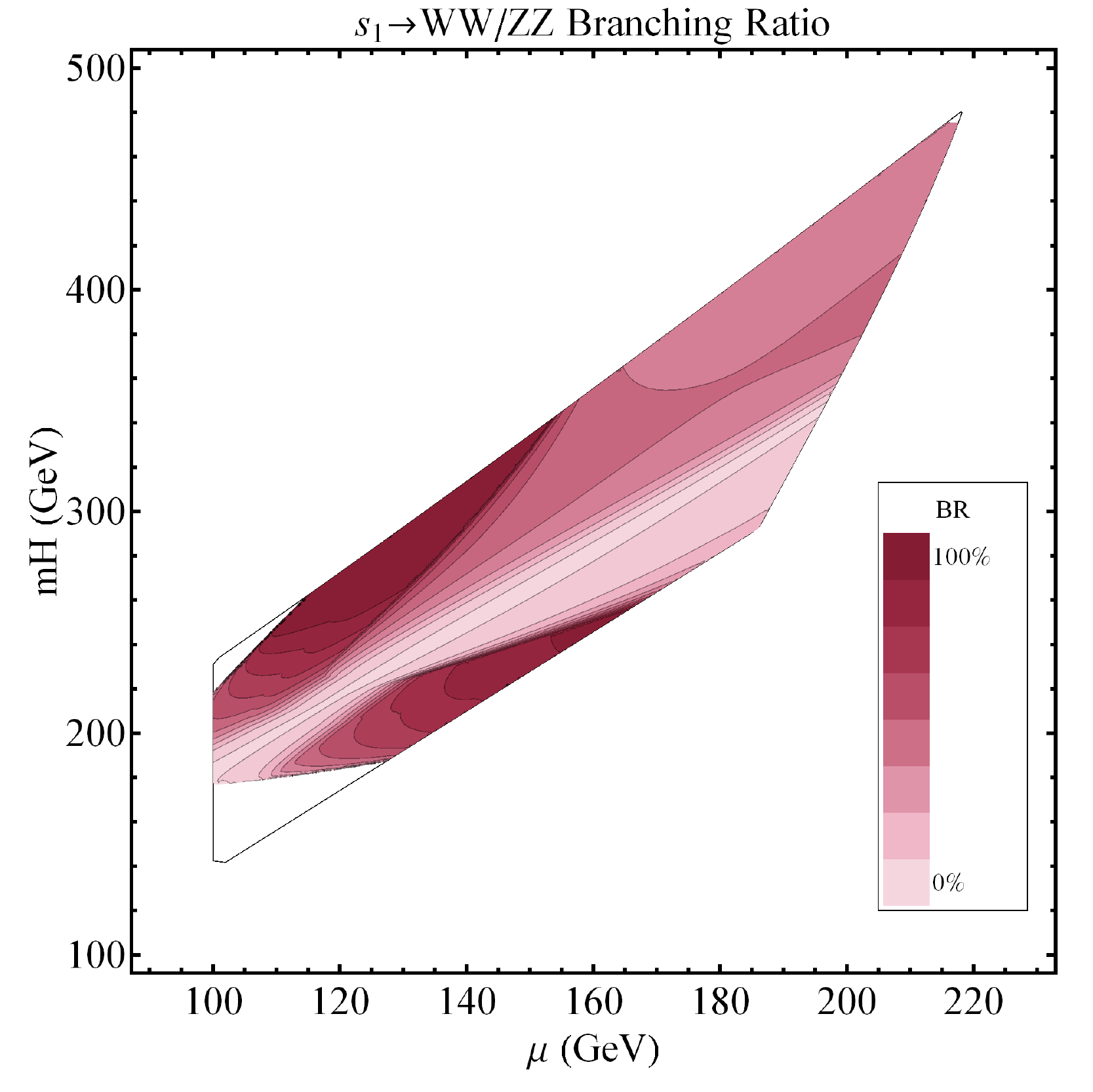}\\
  \end{tabular}
  \includegraphics[width=.36\textwidth]{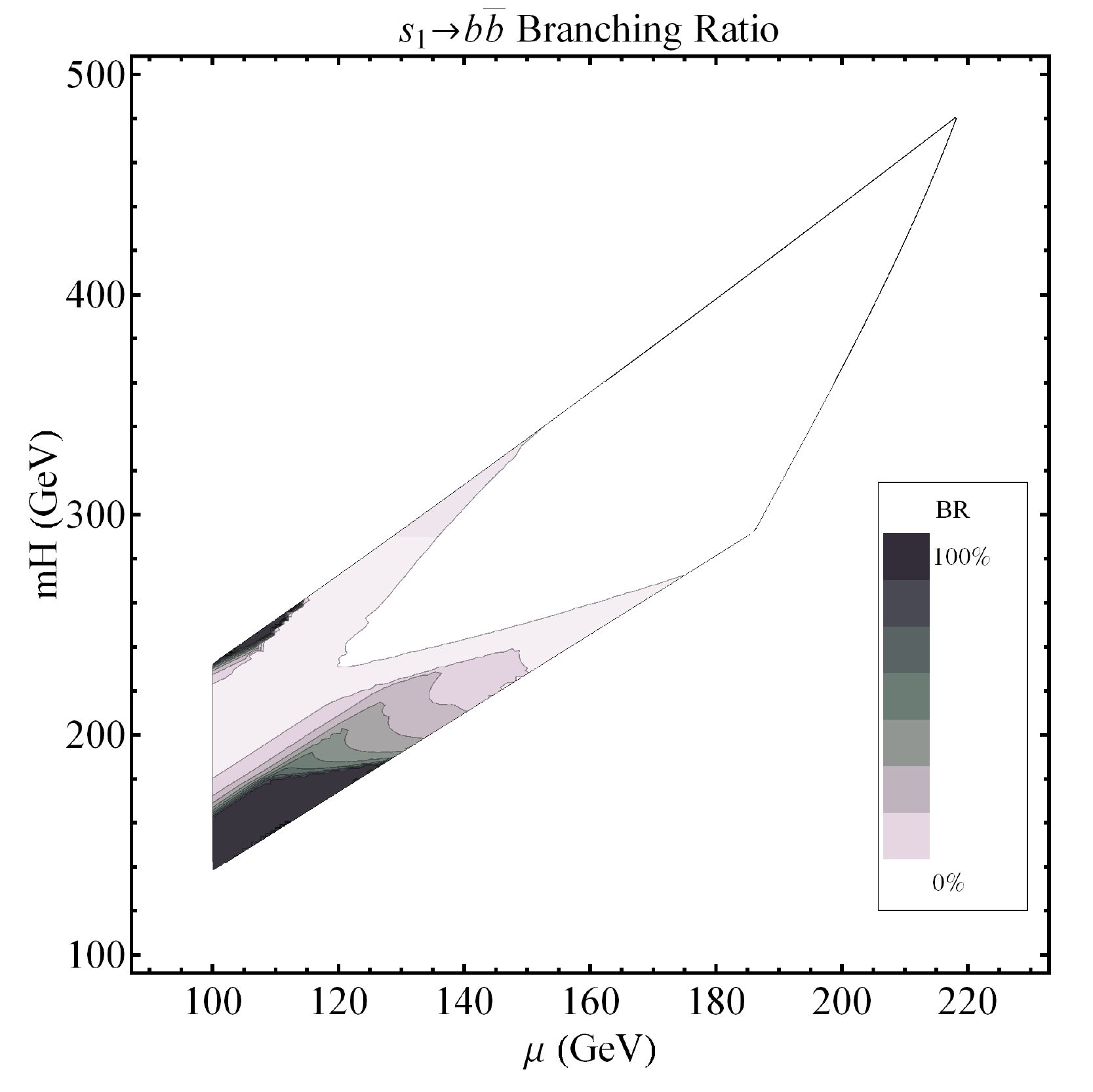}
 \end{center}
\caption{{\small Branching Ratios for the case $k=-0.2$, $\tan\beta=1.5$ and $\lambda=2$. The plots are almost symmetric under
$\mu \rightarrow -\mu$ except for the channel $s_1 \rightarrow \chi_1\chi_1$ which is closed (since in the $\mu<0$ region
the neutralino mass is higher than in the $\mu>0$ region).}}
\label{fig:BR1}
\end{figure}

\begin{figure}[H!t]
 \begin{center}
  \begin{tabular}{cc}
   \includegraphics[width=.4\textwidth]{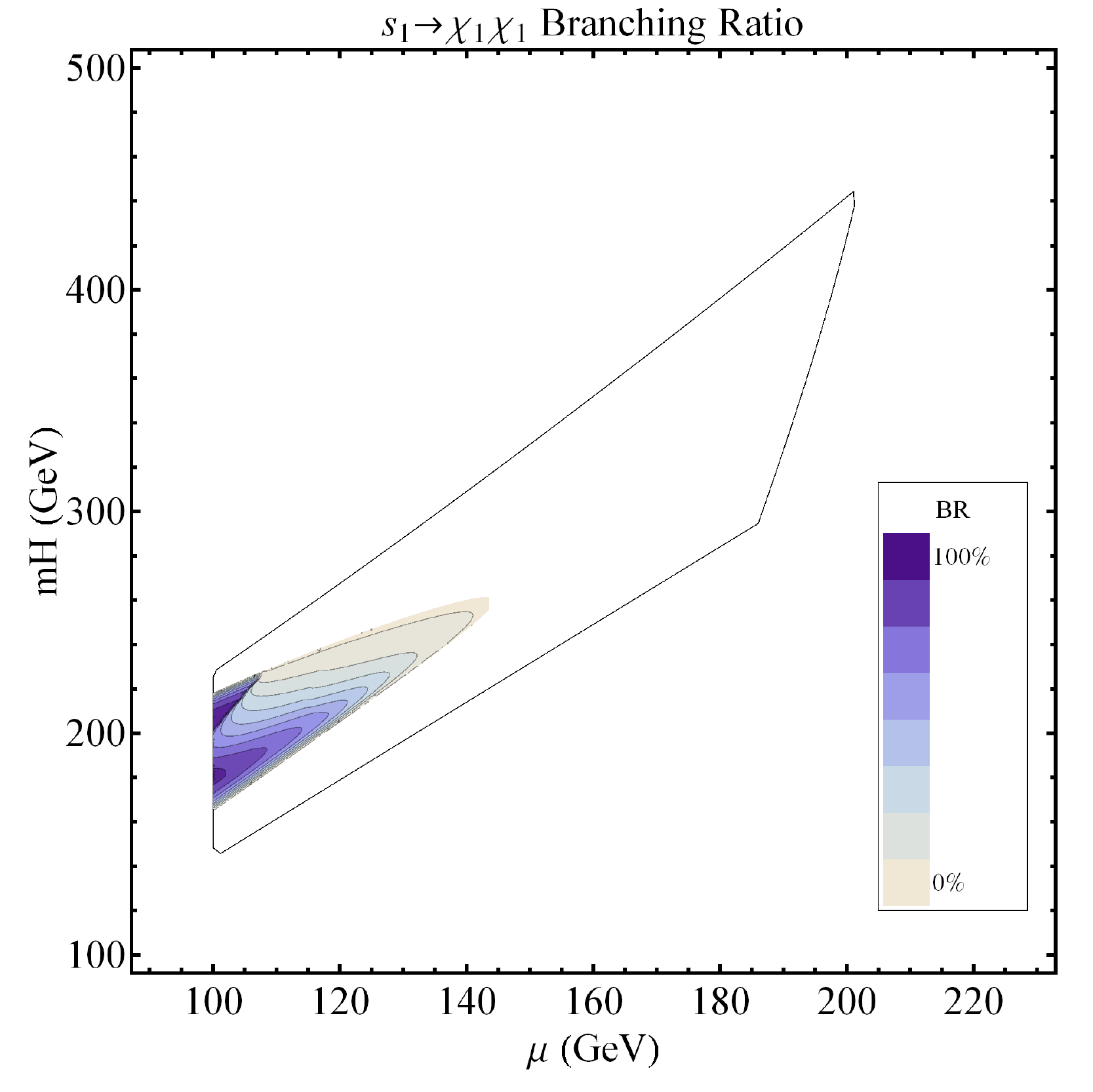} &
   \includegraphics[width=.4\textwidth]{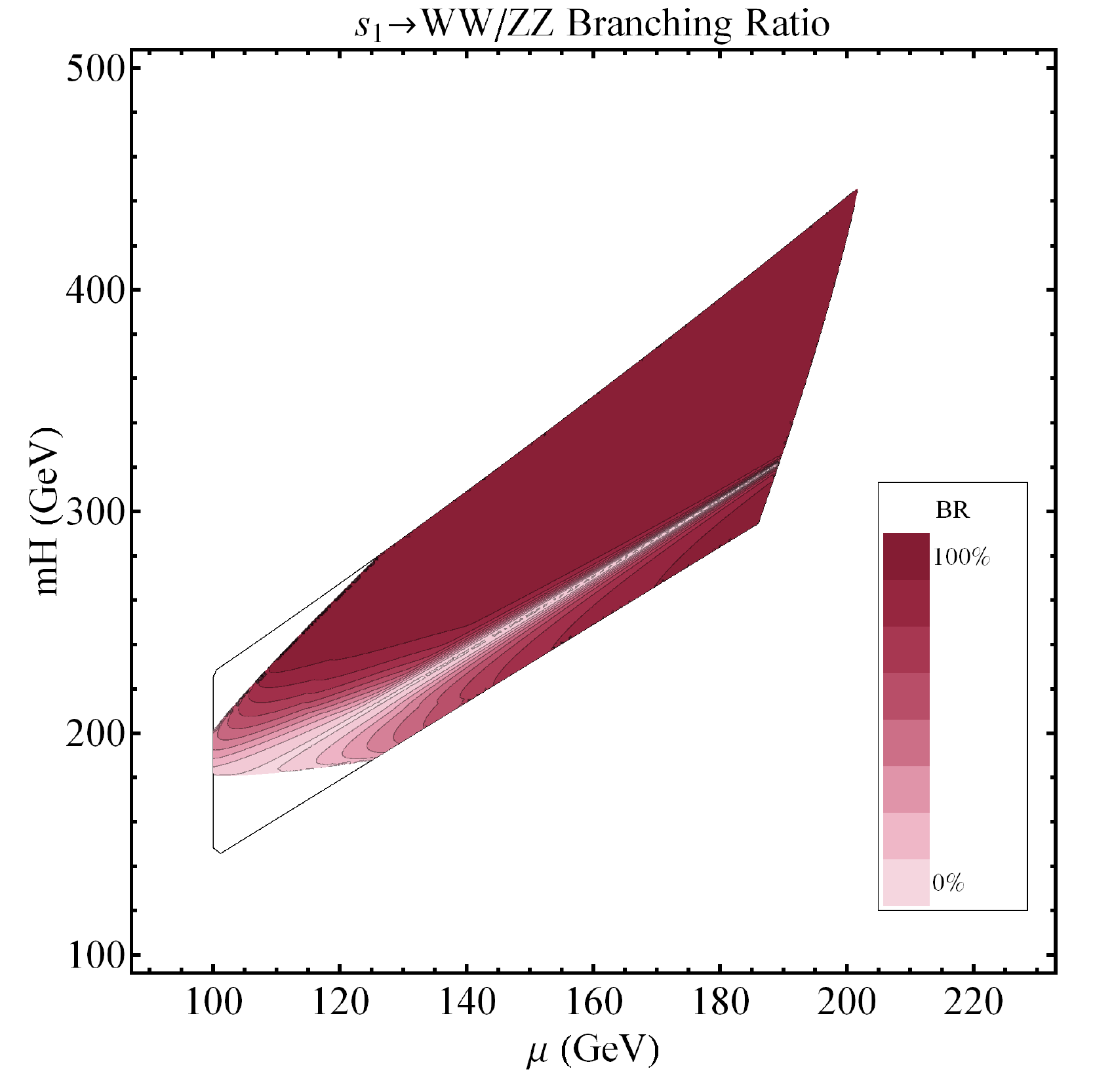}\\
  \end{tabular}
\includegraphics[width=.4\textwidth]{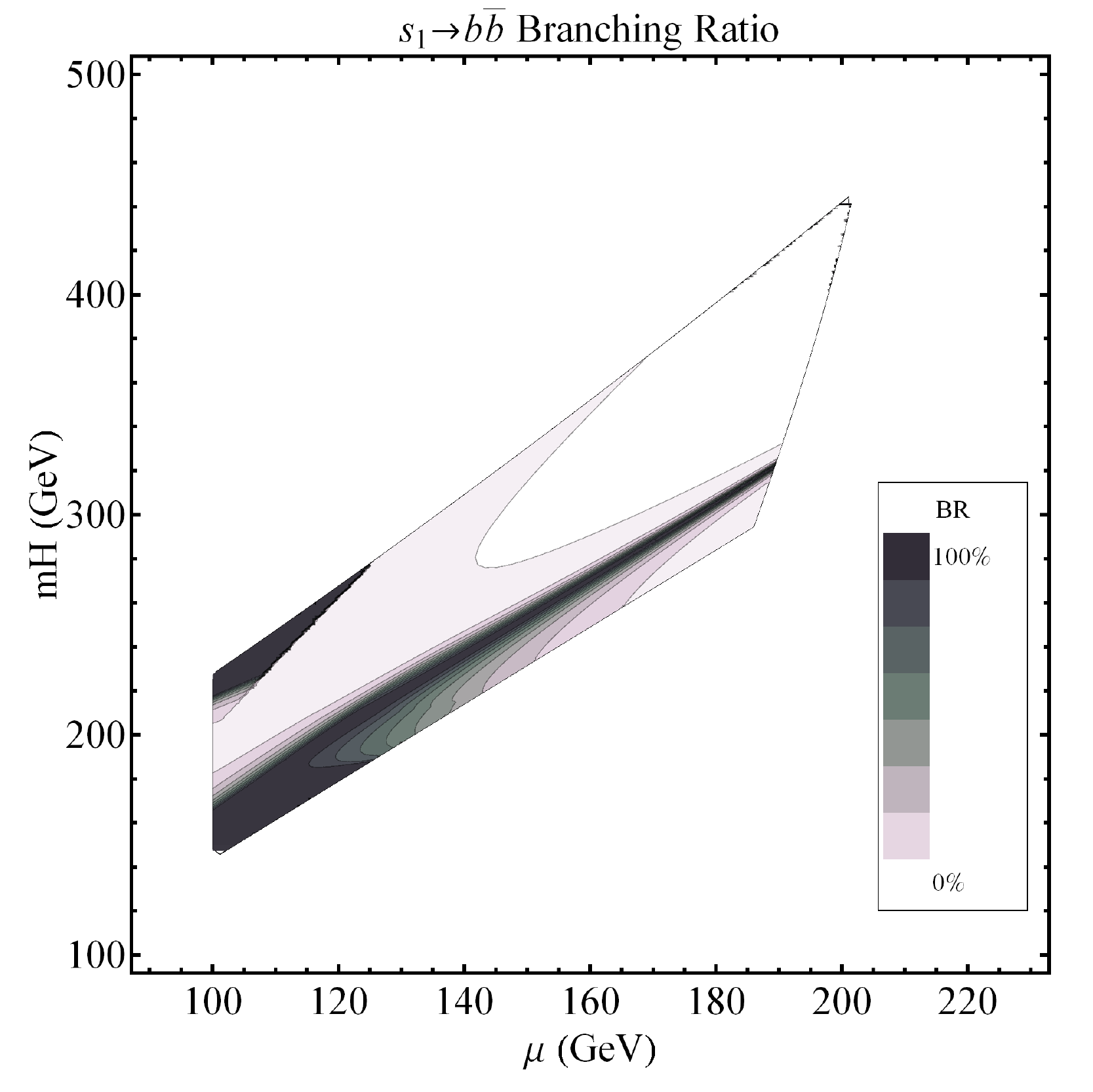}
 \end{center}
\caption{Branching Ratios for the case $k=-0.6$, $\tan\beta=1.5$ and $\lambda=2$. The plots are almost symmetric under
$\mu \rightarrow -\mu$ except for the channel $s_1 \rightarrow \chi_1\chi_1$ which is closed (since in the $\mu<0$ region
the neutralino mass is higher than in the $\mu>0$ region).}
\label{fig:BR2}
\end{figure}
\subsection{Production}
The most relevant feature of the $\lambda$SUSY model is the enhanced Higgs boson mass. Despite this, it is clear that
the lightest scalar in general is not simply a heavy standard Higgs particle (where with standard we mean that it has
the usual couplings to the Standard Model fermions and gauge bosons) but, containing a Singlet component, it can have
very different production rate and Branching Ratios.\\
Concerning the production rate, the relevant quantity is the squared ratio between the couplings of the lightest
scalar to the up-type quarks and to the gauge bosons with respect to the analogous couplings of the standard Higgs particle
with the same mass; these are given by
\begin{eqnarray}\label{eq:xitts1}
\xi_{tts_i} &=& \left(\frac{\sin\beta U_{i2}^*-\cos\beta U_{i1}^*}{s_\beta}\right)^2\\
\xi_{WWs_i} &=& U_{i2}^2\label{eq:xiWWs1}
\end{eqnarray}
where the mixing matrix $U$ is defined by
\begin{equation}
\mathrm{diag}(m_{s_1}, m_{s_2}, m_{s_3}) = U
\begin{pmatrix}
s_\beta & c_\beta & 0
\cr
-c_\beta & s_\beta & 0
\cr
0 & 0 & 1
\end{pmatrix}
M_S^2
\begin{pmatrix}
s_\beta & -c_\beta & 0
\cr
c_\beta & s_\beta & 0
\cr
0 & 0 & 1
\end{pmatrix}  U^{-1}
\end{equation}
In Fig. \ref{fig:percS} we show the Singlet component in the
lightest scalar $s_1$ (upper panel), $\xi_{tts_1}$ (middle panel) and $\xi_{WWs_1}$ (bottom panel). It is clear that in the allowed
 region $s_1$ is never mostly a Singlet. Despite this, its production cross section via gluon-gluon fusion
is always reduced by at least $30\%$\footnote{We explicitly checked that the inclusion of the stop loop does not change this
conclusion.}, and the same is true for the production via vector boson fusion. This is due to the fact
that, in the region we are considering, the two terms at the numerator of eq. (\ref{eq:xitts1}) have same size and
same sign so that there is a partial cancellation. Also in the gauge bosons case we obtain the same
range of values, since the vectors couple only to the combination $c_\beta h_1+ s_\beta h_2$ \cite{Miller:2003ay} which in
general does not correspond to $s_1$.
It is clear that this reduction in the production of the lightest scalar makes harder its detection in the first phase
of the the LHC, also in the
favourable case in which the Branching Ratio in vector bosons is comparable to the one of the standard Higgs boson
with the same mass as $s_1$.
\begin{center}
\begin{table}
\centering
 \begin{tabular}{c|ccccc}
$k=-0.2$ & $\mu$ (GeV) & $m_H$ (GeV) & $m_{s_1}$ (GeV) & $m_{A_1}$ (GeV) & $m_{\chi_1}$ (GeV)\\
\hline
a & 180& 340 & 252 & 103 & 130\\
b & 105 & 180 & 163 & 95 & 77 \\
c & 130 & 200 & 173 & 108 & 96\\
$k=-0.6$ & & & & & \\
\hline
d & 105& 180& 160& 166 & 78\\
e & 160 & 280 & 232 & 195 & 120\\
\end{tabular}
\label{Tab:masses}
\caption{Mass spectrum for some points in parameter space. The corresponding Branching Ratio is reported in Table \ref{Tab:BRS}.
The other parameters are $\tan\beta=1.5$, $\lambda=2$, $M_1=200$ GeV and $M_2=2$ TeV.}
\end{table}
\end{center}

\subsection{Decays}
It is clear that, given the peculiar mass spectrum, in addition to the usual decay channels,
the lightest scalar can also decay into a pair of
lightest pseudoscalars $A_1 A_1$, into $A_1 Z$ and into a pair of neutralinos $\chi_1 \chi_1$.\\
 Whenever the decay mode
$s_1 \rightarrow A_1 A_1$ is open, we expect it to dominate
over all the other channels, since its coupling contains a term that grows as $\lambda^2$ coming from the $S-H_1-H_2$ vertex.
This is true with the exception of a region in which, due to a cancellation in the $s_1-A_1-A_1$ coupling, the dominant mode
is $s_1 \rightarrow A_1 Z$.
At the same time, given that the $s_1-\chi_1-\chi_1$ coupling contains a term that grows as $\lambda$,
in general we expect it to be subdominant with respect to $s_1 \rightarrow A_1 A_1$ but dominant over all the other
decay modes. This is confirmed by Figs. \ref{fig:BR1}-\ref{fig:BR2} in which the Branching Ratios for the different open
channels is shown. \\
In the $k=-0.2$ case, $A_1$ and $\chi_1$ are sufficiently light that in most of the parameter space
at least one between the $s_1\rightarrow A_1 A_1$ and $s_1 \rightarrow \chi_1 \chi_1$ decay channels is open, and they
dominate over all the other decay modes. The situation is almost symmetric under the exchange $\mu \rightarrow -\mu$ with
the exception of the $\chi_1\chi_1$ decay channel, closed due to the higher neutralino mass in the $\mu<0$ region.\\
For $k=-0.6$ the mass of $A_1$ is usually higher so that the dominant decay channel into a pair of pseudoscalars is closed.
In this case in most of the allowed parameter space the
lightest Higgs boson has the same Branching Ratios into $WW$, $ZZ$ and $b\bar{b}$ as the standard Higgs particle, with the exception of a small region in
which the decay into neutralinos dominates. The previous consideration about the $\mu<0$ region apply also in this case.
We collect in Table \ref{Tab:masses}-\ref{Tab:BRS} the mass spectrum and the Branching Ratios for some representative points of
the parameter space.\\
A comment about the $A=G$ choice is in order. We
checked that, relaxing this condition, the $k>0$ case is again almost completely excluded, while in the $k<0$ case there can
be some modification of the allowed region (although never dramatic), but as in the $A=G$ case when the $A_1A_1$ channel
is open it dominates almost everywhere over the others, while when it is closed in a consistent portion of the parameter space
the dominant decay is the one in gauge bosons.
\begin{center}
\begin{table}
\centering
 \begin{tabular}{c|cccccc}
$k=-0.2$ & BR$(A_1A_1)$ & BR$(Z A_1)$ & BR$( \chi_1\chi_1)$ &
BR$(ZZ+ WW)$ & BR$(b\bar{b})$ & $\Gamma_{tot}$ (GeV)\\
\hline
a & 0.51 & 0.09& 0 & 0.38 & 0 & 7\\
b & 0 & 0 & 0.7 & 0.05 & 0.24 & 0.04\\
c & 0 & 0 & 0 & 0.69 & 0.31 & 0.03\\
$k=-0.6$ & & & & & &\\
\hline
d & 0 & 0 & 0.57 & 0 & 0.43 & 0.03\\
e & 0 & 0 & 0 & 0.95 & 0.05 & 0.3\\
\end{tabular}
\label{Tab:BRS}
\caption{Branching Ratios of the lightest scalar for the same points in parameter space considered in Table \ref{Tab:masses}.
The other parameters are fixed as $\tan\beta=1.5$, $\lambda=2$, $M_1=200$ GeV and $M_2=2$ TeV.}
\end{table}
\end{center}
\section{Conclusions}
With the LHC machine taking data and with the promising projection of a Higgs boson discovery potential up to a mass of
$600$ GeV with an integrated luminosity of $5 fb^{-1}$, it is interesting to consider what happens in models in which
the Higgs boson is non standard. In this work we focused on the $\lambda$SUSY framework, in which the Higgs boson
mass is raised already at tree level in order to satisfy the LEP bound without relying on radiative corrections.
For definitiveness, we studied the particular case of a scale invariant superpotential, in contrast to was what done in \cite{lambdaSUSY}.\\
The difference between the two cases is evident: while the lightest scalar in \cite{lambdaSUSY} is essentially an
heavier standard Higgs boson, since both the production rate and the Branching Ratio into gauge bosons are approximately
the standard ones, the one in the scale invariant case have a very different behaviour, especially in the production rate.
Indeed, while there are regions in parameter space in which the decays are dominated by those into gauge bosons, the
production rate from Gluon Gluon fusion is always reduced at least of the $30\%$. In addition, we saw that whenever
the $s_1\rightarrow \chi_1\chi_1$ and $s_1\rightarrow A_1 A_1$ decay modes are open they dominate over all the other channels.
 While in the first case we have to deal with an invisible decay mode, the Higgs search of the lightest scalar via the cascade
$s_1 \rightarrow A_1 A_1 \rightarrow 4j$ seems to be viable with an integrated luminosity of $100 fb^{-1}$
\cite{Kaplan:2011vf}, unfortunately surely out of the reach of the first LHC run.

\section*{Acknowledgments}

We thank Riccardo Barbieri for important suggestions and for reading the manuscript, as well as
Roberto Franceschini and Stefania Gori
for discussions.
This work is supported in part by the European Programme ``Unification in the LHC Era",  contract PITN-GA-2009-237920 (UNILHC).


\begin{thebibliography}{99}
\bibitem{CMS}
See the CMS Higgs Physics Results webpage, [\href{https://twiki.cern.ch/twiki/bin/view/CMSPublic/PhysicsResultsHIG}{twiki.cern.ch}]






\bibitem{lambdaSUSY}
  R.~Barbieri, L.~J.~Hall, Y.~Nomura, V.~S.~Rychkov,
  Phys.\ Rev.\  {\bf D75 } (2007)  035007
    [\hhref{hep-ph/0607332}],
L.~Cavicchia, R.~Franceschini, V.~S.~Rychkov,
  Phys.\ Rev.\  {\bf D77 } (2008)  055006
  [\hhref{0710.5750}],
  R.~Franceschini, S.~Gori,
  JHEP {\bf 1105 } (2011)  084.
  [\hhref{1005.1070}].



\bibitem{LEPHIGGS}
 R.~Barate {\it et al.} [LEP Working Group for Higgs boson
searches and ALEPH and DELPHI and L3 and OPAL Collaborations],
  Phys.\ Lett.\ B\ {\bf 565} (2003) 61.
  [\hhref{hep-ex/0306033}].


\bibitem{Barbieri:2010pd}
  R.~Barbieri, E.~Bertuzzo, M.~Farina, P.~Lodone, D.~Pappadopulo,
  JHEP {\bf 1008 } (2010)  024
  [\hhref{1004.2256}].



\bibitem{NMSSMreview}
  U.~Ellwanger, C.~Hugonie, A.~M.~Teixeira,
  Phys.\ Rept.\  {\bf 496 } (2010)  1-77
  [\hhref{0910.1785}],
 M.~Maniatis,
  Int.\ J.\ Mod.\ Phys.\  {\bf A25 } (2010)  3505-3602
  [\hhref{0906.0777}].

\bibitem{Miller:2003ay}
  D.~J.~Miller, 2, R.~Nevzorov, P.~M.~Zerwas,
  Nucl.\ Phys.\  {\bf B681 } (2004)  3-30
    [\hhref{hep-ph/0304049}].


\bibitem{Ellwanger:2010nf}
  U.~Ellwanger,
  Phys.\ Lett.\  {\bf B698 } (2011)  293-296
  [\hhref{1012.1201}].

\bibitem{Cao:2011pg}
  J.~Cao, Z.~Heng, T.~Liu, J.~M.~Yang,  [\hhref{1103.0631}.


\bibitem{Kaplan:2011vf}
  D.~E.~Kaplan, M.~McEvoy,   [\hhref{1102.0704}].



\bibitem{Almarashi:2010jm}
  M.~Almarashi, S.~Moretti,
  Eur.\ Phys.\ J.\  {\bf C71 } (2011)  1618
  [\hhref{1011.6547}].

\bibitem{Mahmoudi:2010xp}
  F.~Mahmoudi, J.~Rathsman, O.~Stal, L.~Zeune,
  Eur.\ Phys.\ J.\  {\bf C71 } (2011)  1608
   [\hhref{1012.4490}].

\bibitem{Domingo:2011rn}
  F.~Domingo, U.~Ellwanger,  [\hhref{1105.1722}].










\bibitem{Romao:1986jy}
  J.~C.~Romao,
  Phys.\ Lett.\  {\bf B173 } (1986)  309.


\bibitem{Cerdeno:2004xw}
  D.~G.~Cerdeno, C.~Hugonie, D.~E.~Lopez-Fogliani, C.~Munoz, A.~M.~Teixeira,
  JHEP {\bf 0412 } (2004)  048
  [\hhref{hep-ph/0408102}].







\bibitem{Kanehata:2011ei}
  Y.~Kanehata, T.~Kobayashi, Y.~Konishi, O.~Seto, T.~Shimomura,
[\hhref{1103.5109}].





\bibitem{Masses}
  We checked explicitly that our mass matrices are consistent with those in
  J.~R.~Ellis, J.~F.~Gunion, H.~E.~Haber, L.~Roszkowski, F.~Zwirner,
  Phys.\ Rev.\  {\bf D39 } (1989)  844;
  D.~J.~Miller, 2, R.~Nevzorov, P.~M.~Zerwas,
  Nucl.\ Phys.\  {\bf B681 } (2004)  3-30
  [\hhref{hep-ph/0304049}].

\bibitem{Nakamura:2010zzi}
  K.~Nakamura {\it et al.} [ Particle Data Group Collaboration ],
  J.\ Phys.\ G {\bf G37 } (2010)  075021.


\bibitem{Gambino:2001ew}
  P.~Gambino, M.~Misiak,
  Nucl.\ Phys.\  {\bf B611 } (2001)  338-366.
  [\hhref{hep-ph/0104034}].



\end{thebibliography}
\end{document}